\documentclass[iop]{emulateapj}
\usepackage{apjfonts}
\usepackage{natbib}
\usepackage{amsfonts}
\usepackage{amssymb,amsmath}
\usepackage[usenames,dvipsnames]{color}
\usepackage{verbatim}
\newcommand\msun{M$_\odot$}
\newcommand\transpose{\mathsf{T}}

\newcommand{\mathbi}[1]{\textbf{\em #1}}

\newenvironment{packed_item}{
\begin{itemize}
  \setlength{\itemsep}{1pt}
  \setlength{\parskip}{0pt}
  \setlength{\parsep}{0pt}
}{\end{itemize}}
\newcommand{\epssclone}{\epsscale{1.0}}
\newcommand{\epsscltwo}{\epsscale{1.0}}
\newcommand{\epssclthree}{\epsscale{1.0}}

\shorttitle{}
\shortauthors{}

\begin{document}

\title{A GLOBAL TURBULENCE MODLE FOR NEUTRINO-DRIVEN CONVECTION IN
  CORE-COLLAPSE SUPERNOVAE}

\author{Jeremiah W. Murphy\altaffilmark{1,2}}
\author{Casey Meakin\altaffilmark{3,4}}

\altaffiltext{1}{Astronomy Department, The University of Washington
  Seattle, WA 98195, USA; jmurphy@astro.washington.edu}
\altaffiltext{2}{NSF Astronomy and Astrophysics Fellow}
\altaffiltext{3}{Theoretical Division, Los Alamos National Laboratory,
  Los Alamos, NM 87545, USA}
\altaffiltext{4}{Steward Observatory, University of Arizona, Tucson,
  AZ 85721, USA}

\begin{abstract}
Simulations of core-collapse supernovae (CCSNe) result in
successful explosions once the neutrino luminosity exceeds a critical
curve, and recent simulations indicate that turbulence further enables
explosion by reducing this critical neutrino luminosity.  We propose a
theoretical framework to derive this result and take the first steps
by deriving the governing mean-field equations.  Using Reynolds
decomposition, we decompose flow variables into background and
turbulent flows and derive self-consistent averaged equations for
their evolution.  As basic requirements for the CCSN problem, these
equations naturally incorporate steady-state accretion, neutrino
heating and cooling, non-zero entropy gradients, and turbulence terms
associated with buoyant driving, redistribution, and dissipation.
Furthermore, analysis of two-dimensional (2D) CCSN simulations validate these
Reynolds-averaged equations, and we show that the physics of
turbulence entirely accounts for the differences between 1D and 2D CCSN
simulations.  As a prelude to deriving the reduction in the critical
luminosity, we identify the turbulent terms that most influence the
conditions for explosion.  Generically, turbulence equations require
closure models, but these closure models depend upon the macroscopic
properties of the flow.  To derive a closure model that is appropriate
for CCSNe, we cull the literature for relevant closure
models and compare each with 2D simulations.  These models employ local
closure approximations and fail to reproduce the global properties of
neutrino-driven turbulence. Motivated by the generic failure of these
local models, we propose an original model for turbulence which incorporates global properties of
the flow.  This global model accurately reproduces the turbulence
profiles and evolution of 2D CCSN simulations.

\end{abstract}

\keywords{convection --- hydrodynamics ---
  instabilities --- methods:analytical --- methods: numerical ---
  shock waves --- supernovae: general --- turbulence}

\section{Introduction}
\label{introduction}

\par Though our understanding of the core-collapse supernova (CCSN)
explosion mechanism remains incomplete, recent simulations indicate
that it is likely to involve multi-dimensional effects.  In fact, in
all proposed mechanisms neutrino-driven convection plays an important,
if not vital, role.  Motivated by these results, we present a
theoretical framework to investigate the role of turbulence in
launching successful explosions. Furthermore, we lay down the foundation
for this framework by deriving self-consistent steady-state equations
for the background and turbulent flows. 

The fundamental problem of CCSN theory is to determine how the stalled
shock transitions into a dynamic explosion.  Within a few milliseconds
after bounce, the core-bounce shock wave stalls into an accretion shock \citep{mazurek82,bruenn85,bruenn89}.  Unchecked, continued accretion through the shock
would form a black hole \citep{oconnor11}.  However, the
preponderance of observed neutron stars \citep{lorimer06} and
supernova (SN)
explosions \citep{li2010} dictates that the stalled shock is revived
into an explosion most of the time.

\par For more than two decades, the favored mechanism for core
collapse supernovae has been the delayed-neutrino mechanism \citep{bethe85}.  In this model, a neutrino luminosity of several
times 10$^{52}$ erg/s cools the protoneutron star and heats the region
below the shock.  Under the correct conditions, this heating by
neutrinos can revive the shock and produce an explosion.
Unfortunately, except for the least massive stars which undergo core
collapse, most of the detailed 1D neutrino-transport-hydrodynamic
simulations do not produce solutions containing explosions \citep{liebendorfer01a, liebendorfer01b, rampp02, buras03, thompson03, liebendorfer05b,kitaura06}.

\par While most 1D simulations fail to produce explosive
  solutions, recent 2D simulations show promising trends. These
simulations capture multi-dimensional instabilities that aide the neutrino mechanism in driving explosions \citep{marek07,murphy08b,scheck08,marek09b,fernandez09,suwa10,nordhaus10}. These instabilities include neutrino-driven convection \citep{burrows95,janka95,murphy08b,nordhaus10} and the standing accretion shock instability (SASI) \citep{blondin03,ohnishi06,burrows06,marek07,marek09b,foglizzo00,foglizzo09}.  Besides these two instabilities, other multi-dimensional processes may revive the stalled shock, including magnetohydrodynamic (MHD) jets \citep{burrows07c,dessart08a} and acoustic power derived from asymmetric accretion and an oscillating PNS \citep{burrows06,burrows07a}.  Though the prevalence of these last two processes is still debated \citep{dessart08a,weinberg08}, two points remain clear: (1) the solution to the CCSN problem is likely to depend on multi-dimensional effects, and (2) in all proposed mechanisms, neutrinos and turbulence play an important, if not central, role. Hence, whatever the mechanism, it is important to understand the role of neutrinos and turbulence.

\par \citet{burrows93} proposed that if a critical neutrino luminosity
is exceeded for a given mass accretion rate, the neutrino
mechanism succeeds.  These authors developed a steady-state
model for an accretion shock in the presence of a parametrized
neutrino heating and cooling profile.   The two most important
parameters of the steady-state model are the neutrino luminosity
$L_{\nu_e}$ and mass accretion rate $\dot{M}$.  They
found no steady-state solutions for luminosities above a critical
curve $L_{\nu_e,crit} = f(\dot{M})$, and interpreted this curve as
separating steady-state (or failed supernova) solutions from explosive
solutions. However, this work did not prove that the solutions above
the critical curve are in fact explosive, nor did they consider
multidimensional effects.

\par Using a similar neutrino parametrization as in the
\citet{burrows93} work, \citet{murphy08b} showed in 1D and 2D
simulations that the solutions above the critical luminosity are in
fact explosive. Moreover, they found that the critical luminosity in
the 2D simulations is $\sim$70\% of that in 1D for a given $\dot{M}$.
Additional investigations by \citet{nordhaus10} show that the critical
luminosity is even further reduced in 3D.
These results suggest that the critical luminosity is a useful
theoretical framework for describing the conditions for successful
explosions \citep{burrows93,murphy08b}.

Initial investigations by \citet{murphy08b} also suggest that the
reduction in the critical luminosity is caused by turbulence.  An
alternative but related condition for explosion is a comparison of the
advection and heating timescales
\citep{thompson00,janka01,thompson03,murphy08b}.  The heating
timescale is the time it takes to significantly heat a parcel of
matter, and the advection timescale is the time to advect through this
region.  If the advection timescale is long compared to the heating
timescale, then explosion ensues.  In 1D, as matter accretes onto the
PNS, it is limited to advect through the gain region with one short
timescale.  In 2D, convective motions increase the dwell time, which
leads to more heating for the same neutrino luminosity and a lower
critical luminosity\citep{murphy08b}. \citet{pejcha11} have recently
challenged this explanation and suggest that rather than increasing the
heating, turbulence acts to reduce the cooling.  Regardless, the
simulations show that the critical luminosity is lower in the presence
of convection.

\par These results suggest that a theory for successful explosions
requires a theoretical framework for turbulence and its influence   on
the critical luminosity.  In this paper, we develop the foundation for
such a framework.
Recent developments in turbulence theory have led to accurate
turbulence models, and in this paper we use similar strategies to
develop a turbulence model appropriate for CCSNe.  Such a turbulence
model can then be incorporated into steady-state accretion models to
derive reduced critical luminosities for explosion, as well as used in
1D radiation-hydrodynamic (RHD) simulations to expedite systematic
studies of core-collapse physics.  In the present paper, we develop a
turbulence model which captures the salient features  of 2D core
collapse convection,  but eventually the model must be calibrated
against 3D simulations. 

\par To develop a turbulence model appropriate for core-collapse
turbulence, we use a general and fully self-consistent approach called
Reynolds decomposition \citep{plate97,pope00,launder02}.  The first
step in this approach is to decompose the flow variables into averaged
and fluctuating components.  Evolution equations for the mean flow variables are then
developed by writing the conservation laws in terms of these mean and
fluctuating components and then averaging.  The resulting evolution
equations for the mean fields contain terms which involve both the
mean fields as well as correlations of fluctuating components. These correlations
represent the action of turbulence and include the Reynolds stress,
turbulent kinetic energy, turbulent enthalpy flux, and higher order
correlations.  These evolution equations are self-consistent and
naturally include the effects of a background flow, which is important
in core collapse.  Unfortunately, this procedure always produces
evolution equations which depend on a correlation of higher order than
the evolution equations.  Therefore, in order to develop a closed
system of equations, the highest order correlations must be
modeled in terms of the lower order correlations and mean fields.
Furthermore, closure depends upon the macroscopic flow itself, so
there is no unique closure for turbulence.  This is the infamous
closure problem of turbulence.

\par Fortunately, there is a small class of turbulent flows  (e.g., shear, buoyancy driven, etc.) and closure models have been developed that  work well for each type under a range of conditions \citep{turner73,plate97,pope00,launder02,wilcox06}. The general strategy to find closure relations involves an interplay between theory, observations, and numerical simulations \citep{launder02}. First, terms in the mean-field equations are compared to either observations or numerical simulations.  Approximations are then proposed for the higher order correlation terms that satisfy the observations/simulations and provide closure.  This approach has been used successfully for geophysical flows \citep{launder02} and is now being applied to stellar structure calculations \citep{garaud10, meakin07b}.

Following on these successes, we use this strategy
  to develop a turbulence model for the core-collapse problem in which
  buoyancy and a background accretion flow dominate.  
  In \S~\ref{section:reynoldsdecomp}, we use Reynolds
decomposition to formally derive the averaged background and turbulence equations and identify
terms that are important for neutrino-driven convection.  
Using 2D simulations, we examine in \S~\ref{section:compare2d} the turbulent properties of
neutrino-driven convection and show that the turbulence equations
which we derive in \S~\ref{section:reynoldsdecomp} are consistent with the simulated flows.
Finding solutions to
the mean-field equations requires a closure model.  
Therefore, in \S~\ref{section:models} we present several models representative
of the literature.  However, these fail to reproduce the global
profiles of neutrino-driven convection, leading us to develop a novel
global model.
In \S
\ref{section:comparemodels}, we compare the results of the turbulence
models (\S~\ref{section:models}) with the results of 2D simulations
and conclude that our global model is the only model to reproduce the
global properties of neutrino-driven convection.
In \S~\ref{section:conditions}, using the mean-field equations and 2D
simulations, we investigate the effects of turbulence on the
conditions for successful explosions.
Finally, in \S\ref{section:conclusion} we summarize our findings, and
motivate the need for a similar analysis using 3D simulation data.

\section{Background and Turbulence Equations: Reynolds Decomposition}
\label{section:reynoldsdecomp}

The first step in understanding the effects of turbulence is to derive
the governing steady-state equations.  Therefore, in this section, we
use Reynolds decomposition to derive exact equations for the
steady-state background and turbulent flows. 

Historically, turbulence modelers have used two approaches to derive
the models.  In one, ad hoc equations are suggested to model
the important turbulence physics.  These are often of practical use,
but the underlying assumptions often make these of limited use.
Mixing length theory (MLT) is one
such approach (see \S \ref{section:algebraic} for its limitations) \citep{kippenhahn90}.
In the second approach, one derives self-consistent equations for
turbulence by decomposing the hydrodynamic equations into background
and turbulent parts.  Reynolds decomposition is an example of this
approach.  Though these equations are
exact, they are not complete; they need a model for
closure.  If the ad hoc approaches accurately represent Nature, then
one should be able to derive them by making the appropriate
assumptions in the exact equations.  Therefore, regardless
of the technique employed, starting with the self-consistent equations
enables a better understanding of the assumptions and limitations. In this paper, we pursue both approaches, but in this
section, we use Reynolds decomposition to derive and explore the
self-consistent equations for the background and turbulent
flows.

In Reynolds decomposition, the hydrodynamic equations are decomposed
into background and turbulent flows \citep{pope00}.
Consider a generic flow variable, $\phi$, and
its decomposition into average (background) and fluctuating
(turbulent) components: $\phi = \phi_0 + \phi^{\prime}$.  The
mean-field background of $\phi$, $\left < \phi \right >$, is
obtained by coarse spatial and temporal averages.  The
interval for the averages must be large or long enough to smooth out
short term turbulent fluctuations, but they must not be too large or
long so that interesting spatial or temporal trends in the mean-field
quantities are completely averaged out.  Choosing the scales of the
averaging window is dependent upon the problem, and in this
paper, we define $\langle \rangle$ as averaging over the solid angle in the
spherical coordinate system and over a fraction of the
eddy crossing time of the convective region.  By definition, the
coarse average of $\phi$ is $\left < \phi \right > =
\phi_0$ and the mean-field average of the fluctuation is identically
zero, $\left < \phi^{\prime}\right > =0$. Therefore, first order
moments of turbulent fluctuations are identically zero and only higher order
terms survive.  For example, the average of the velocity fluctuation is
zero, $\left < \mathbi{v}^{\prime} \right >$, but the mean-field of
the second order term, the Reynolds stress $\left < \mathbf{R} \right
> = \left < \mathbi{v}^{\prime} \otimes \mathbi{v}^{\prime} \right >$,
is nonzero.  

\subsection{Averaged Background Equations}
\label{section:averagebackground}

The general equations for mass, momentum, and entropy conservation are
\begin{equation}
\label{eq:masshydro}
\frac{\partial \rho}{\partial t} 
+ \nabla \cdot \left ( \rho \mathbi{u} \right ) 
= 0 \, ,
\end{equation}
\begin{equation}
\label{eq:momentumhydro}
\frac{\partial \left ( \rho \mathbi{u} \right )}{\partial t}
+ \nabla \cdot \left ( \rho \mathbi{u} \otimes \mathbi{u} \right )
= - \nabla P
+ \rho \mathbi{g} \, ,
\end{equation}
and
\begin{equation}
\label{eq:entropyhydro}
T \left ( 
\frac{\partial \left ( \rho s \right )}{\partial t}
+ \nabla \cdot \left ( \rho s \mathbi{u} \right ) 
\right)
= \rho \dot{q} \, .
\end{equation}
In these equations, the density, velocity, pressure, temperature, and
specific entropy are $\rho$, $\mathbi{u}$, $P$, $T$, and $s$.
$\mathbi{g}$ is the gravitational acceleration, and $\dot{q}$ is the
specific local heating and/or cooling rate.

After Reynolds decomposition, averaging, and assuming steady state,
the hydrodynamics equations,
eqs.~(\ref{eq:masshydro}-\ref{eq:entropyhydro}), become
\begin{equation}
\label{eq:mass}
\nabla \cdot (\rho_0 \mathbi{v} + \left < \rho^{\prime} \mathbi{v}^{\prime} \right >)
= 0 \, ,
\end{equation}
\begin{equation}
\label{eq:momentum}
\left < \rho \mathbi{u} \right > \cdot \nabla \mathbi{v} = 
-\nabla P_0
+\rho_0 \mathbi{g}
- \nabla \cdot \left < \rho \mathbf{R} \right > \,
\end{equation}
and 
\begin{equation}
\label{eq:entropy}
\left < \rho \mathbi{u} \right > \cdot \nabla s_0 = 
\left < \frac{\rho \dot{q}}{T} \right >
+ \frac{\rho_0 \epsilon}{T_0}
- \nabla \cdot \left < \mathbi{F}_s \right > \, .
\end{equation}
For all quantities, the turbulent perturbations are denoted by a superscript ${\prime}$, and, with the
exception of velocity, the background flow is denoted by subscript
$0$.  The background velocity is $\mathbi{v}$,
and the perturbed velocity is $\mathbi{v}^{\prime}$.

Equations \ref{eq:mass}-\ref{eq:entropy} are very similar to the usual
steady-state equations of hydrodynamics, but the last term in all
three equations add new turbulence physics.  Conservation of mass flux
is split between the background and the
turbulence, $\left < \rho^{\prime} \mathbi{v}^{\prime} \right >$.  In
the momentum equation, the extra force due to turbulence is the divergence of the Reynolds stress,
$\mathbf{R} = \mathbi{v}^{\prime} \otimes \mathbi{v}^{\prime}$.  The entropy equation has two new terms; the
divergence of the entropy flux, $\mathbi{F}_s = \rho
s^{\prime} \mathbi{v}^{\prime}$, represents entropy redistribution by turbulence, and
$\epsilon$ represents heat due to turbulence dissipation.  For
isotropic turbulence, the divergence of $\mathbf{R}$ can be re-cast as the gradient of turbulent
pressure: i.e. $-(2/3)\nabla (\rho K)$, where $K = \mathbi{v}^{\prime} \cdot
\mathbi{v}^{\prime}/2$ is the turbulent kinetic energy.  Using
thermodynamic relations, we reduce the number of turbulent
correlations in eqs.~(\ref{eq:mass}-\ref{eq:entropy}) by noting
that $\left < \rho^{\prime} \mathbi{v}^{\prime} \right > =
\eta \mathbi{F}_s$, where $\eta = \beta_T/c_P$, $c_P$ is the specific heat at
constant pressure, and $\beta_T = - (\partial \ln \rho/\partial \ln
T)_{P}$ is the logarithmic derivative of density with respect to
temperature at constant pressure.  

While we consider the convective entropy flux, $\mathbi{F}_s$,
traditionally, astrophysicists have considered the enthalpy flux,
$\mathbi{F}_e = \rho_0 c_P \left < \mathbi{v}^{\prime} T^{\prime}
\right > $ in turbulence models.  The enthalpy flux has units of
energy flux.  Therefore, the enthalpy flux is a natural choice for
stellar structure calculations in which the enthalpy flux and
radiative flux must add to give the total
luminosity of the star.  Because the convective region is
semi-transparent to neutrinos, there is no such constraint in the
core-collapse problem.  Furthermore, since we decompose the entropy
equation, the entropy flux is the most natural flux to consider.  For
the instances that require discussing the enthalpy flux, we express it
in terms of the entropy flux, $\mathbi{F}_e = T_0 \mathbi{F}_s$.

In expressing ($\langle \rho^{\prime} \mathbi{v}^\prime \rangle = \eta
\mathbi{F}_s$, we have eliminated one of
the turbulent correlations , but there still remain three turbulent correlations ($\mathbf{R}$,
$\mathbi{F}_s$, and $\epsilon$), resulting in more unknowns than
equations.  To close these equations, we derive the turbulence
equations in \S~\ref{section:turbulenceequations}).

\subsection{Averaged Equations for Turbulent Correlations}
\label{section:turbulenceequations}

Using the definitions of $\mathbf{R}$ and $\mathbi{F}_s$ and the conservation
equations, we re-derive the evolution equations for the Reynolds
stress ($\mathbf{R}$) and the entropy flux
($\mathbi{F}_s$).  For similar derivations of these equations, see
\citet{canuto93} and \citet{garaud10}.  The convective Reynolds stress equation is
\begin{equation}
\label{eq:reynoldsstress}
\begin{array}{lllr}
\lefteqn{\partial \langle \rho \mathbf{R} \rangle / \partial t
+ \mathbi{v} \cdot \nabla \left < \rho \mathbf{R} \right > 
+ \left < \rho \mathbf{R} \right > \otimes \nabla \cdot \mathbi{v} = } \\
& & + \left < \rho^{\prime} \mathbi{v}^{\prime} \right > \otimes
\mathbi{g} + [\left < \rho^{\prime} \mathbi{v}^{\prime} \right > \otimes
\mathbi{g}]^{\transpose}
& \textrm{Buoyant production} \\
& & - \left < \rho \mathbf{R} \right > \cdot \nabla \mathbi{v} - [\left < \rho \mathbf{R} \right > \cdot \nabla \mathbi{v}]^{\transpose}
& \textrm{Shear production} \\
& & - \langle  \nabla \otimes \mathbi{F}_P +
       [\nabla \otimes \mathbi{F}_P]^{\transpose} \rangle  
& \textrm{Pressure flux}\\
& & + \langle  P^{\prime} \nabla \mathbi{v}^{\prime}  +
       [P^{\prime} \nabla \mathbi{v}^{\prime}]^{\transpose} \rangle  
& \textrm{Pressure strain}\\
& & - \nabla \cdot \left < \rho \mathbi{v}^{\prime} \otimes \mathbf{R} \right > 
& \textrm{Turbulent transport} \\
& & - \rho_0 \boldsymbol{\varepsilon} 
& \textrm{Dissipation} \, , \\
\end{array}
\end{equation}
where the turbulent pressure flux is $\mathbi{F}_P = P^\prime \mathbi{v}^\prime$, the dissipation tensor is $\rho_0 \boldsymbol{\varepsilon} = \mu [ \nabla^2\left < \mathbf{R} \right > -
2 (\nabla \mathbi{v}^{\prime}) \cdot (\nabla \mathbi{v}^{\prime}) ]$,
and $[]^{\transpose}$ is the transpose operator.
In eq. (\ref{eq:reynoldsstress}), we separate the terms on the right-hand-side
into rows to better illustrate their physical relevance.  They
are buoyant and shear production, redistribution by the turbulent
pressure flux, the pressure-strain correlation, turbulent Reynolds
stress transport,
and the turbulent dissipation.  In neutrino-driven convection of
core collapse, buoyancy is the most important turbulent
production.  In terms of driving
turbulence, the shear production term is less important.  However,
this term
and the pressure-strain term are primarily responsible for
redistributing stress among the
components.  For example, gravity acts mostly on the vertical stress
components, but the shear production and pressure-strain terms
redistribute stress to the horizontal components.  Also important in
redistributing stress is the turbulent transport term.  In fact, in the next paragraph, we show that this
term is in effect the divergence of the turbulent kinetic energy flux,
which is very important in vertical kinetic energy transport.  

Taking the trace of the Reynolds stress equation gives 
the convective kinetic energy equation:
\begin{equation}
\label{eq:kinetic}
\begin{array}{llll}
\lefteqn{ \partial \langle \rho K \rangle / \partial t
+\mathbi{v} \cdot \nabla \left < \rho K \right > + \left < \rho K
\right > \nabla \cdot \mathbi{v} = } 
\\
& & & 
+ \left < \rho^{\prime} \mathbi{v}^{\prime} \right > \cdot \mathbi{g}
- \mbox{tr}\left ( \left < \rho \mathbf{R}\right > \cdot \nabla \mathbi{v}\right )
\\
& & & 
- \nabla \cdot \left < \mathbi{F}_{K} \right >
- \nabla \cdot \langle \mathbi{F}_P \rangle  
+ \left < P^{\prime} \nabla \cdot \mathbi{v}^{\prime}  \right >
- \rho_0 \epsilon \, ,
\\
\end{array}
\end{equation}
where $\mbox{tr}()$ is the trace operator, the turbulent dissipation
becomes $\epsilon = \mbox{tr}(\boldsymbol{\varepsilon})/2$, and
$\mathbi{F}_K =  \rho \mathbi{v}^{\prime} K$ is the turbulent
kinetic energy flux.  Once again on the right-hand-side, we have the
familiar terms: buoyancy and shear production, turbulent
redistribution by the turbulent kinetic energy flux, the
divergence of the pressure flux, work done by turbulent pressure, and
turbulent dissipation.

The corresponding equation for convective entropy flux is
\begin{equation}
\label{eq:entropyflux}
\begin{array}{lllr}
\lefteqn{\partial \langle \mathbi{F}_s \rangle / \partial t
+\mathbi{v} \cdot \nabla \left < \mathbi{F}_s \right > + \left
  < \mathbi{F}_s \right> \nabla \cdot \mathbi{v} = } \\
& & + \rho_0 \eta \left < Q \right > \mathbi{g} 
& \textrm{Buoyant production} \\ 
& & -\left < \mathbi{F}_s \right > \cdot \nabla \mathbi{v} 
- \left < \rho \mathbf{R} \cdot \right > \nabla s_0
& \textrm{Gradient production} \\
& & - \left < s^{\prime} \nabla P^{\prime}\right >
& \textrm{Pressure covariance} \\
& & - \nabla \cdot \left < \mathbi{v}^{\prime} \otimes \mathbi{F}_s \right > 
& \textrm{Turbulent transport} \\
& & + \left< \rho \mathbi{v}^{\prime} \dot{q}/T \right >
& \textrm{heat production} \, . \\
\end{array}
\end{equation}
where $Q = {s^{\prime}}^2$ is the variance of the entropy perturbation.
Once again, we separate the terms in eq.~(\ref{eq:entropyflux}) into rows to
highlight their physical significance.  The terms in the entropy flux
equation are analogous to the terms in the Reynolds stress equation.
They are buoyant and gradient production terms, the
pressure-covariance, turbulent transport,
and heat production.  Unlike the Reynolds stress equation, we
find that buoyant production, gradient production, pressure
covariance, and turbulent transport are all equally relevant in
determining the entropy flux.

The first term on the right hand side of eq.~(\ref{eq:entropyflux}) is the
buoyant production.  This term is an important source in
eq.~(\ref{eq:entropyflux}), but it depends upon yet another correlation, the
variance of the entropy perturbation.  The corresponding equation for the entropy variance is
\begin{equation}
\begin{array}{lllr}
\label{eq:entropyvariance}
\lefteqn{\partial \langle \rho Q \rangle / \partial t
+ \nabla \cdot \left ( \rho \mathbi{v} \langle Q \rangle \right ) 
=} \\
& & -2 \left < \mathbi{F}_s \right > \cdot \nabla s_0
& \textrm{Gradient Production} \\
& & + 2 \langle s^{\prime} \dot{Q}/T \rangle 
& \textrm{Heat Production} \\
& & - \nabla \cdot \left < \rho \mathbi{v}^{\prime} Q \right > 
& \textrm{Turbulent Transport} \, . \\
\end{array}
\end{equation}

Equations (\ref{eq:reynoldsstress}-\ref{eq:entropyvariance}) are an
exact set of evolution equations for the 2$^{\rm{nd}}$ order correlations (i.e. Reynolds
stress, entropy flux, and entropy variance).  While these equations
are exact, they are not complete.  Each equation depends upon 3$^{\rm{rd}}$
order correlations, necessitating further evolution equations for the
higher order correlations.  However, it is impossible to close the
turbulence equations in this way, as each set of evolution equations
depends upon yet higher order correlations.  The only solution is to
develop a closure model to relate higher order moments to lower order
moments.

This is analogous to the closure problem in deriving the hydrodynamics
equations.  In the hydrodynamics equations, the equation of state (EOS) is a microphysical
closure model which relates the
pressure (a higher moment) to the density and internal energy.
Because of the vast separation of scale, the EOS depends upon
microphysical processes only and is independent of the
macroscopic hydrodynamical flows.  Hence, as a closure model, the EOS
enables the hydrodynamic equations to be relevant for a wide range of
macroscopic flows.  Conversely, turbulence occupies the full range of
scales from the microscopic to the largest bulk flows.  In some cases,
turbulence is the dominant macroscopic flow.  Consequently, closure is
necessarily dependent upon the macroscopic flow, making it impossible
to derive a generic closure relation for turbulence.

To find solutions to the turbulence equations, we need to construct a
turbulence closure model that is appropriate for core collapse.  The
standard approach is to develop a turbulence closure model for each macroscopic flow.  Fortunately, this task is not as daunting as
it first appears.  Turbulence can be divided into several classes that
are characterized by the driving mechanism (i.e. shear, buoyancy,
magnetic) and closure models have been constructed that are
appropriate for each class.  For core collapse, buoyancy is the
primary driving force, and the rest of this paper is devoted
to finding an appropriate buoyancy closure
model for core-collapse turbulence.

\subsection{Steady-state Reynolds-averaged Equations in Spherical Symmetry}

Assuming a spherically symmetric background, the equations
for the background flow (eqs. \ref{eq:mass}-\ref{eq:entropy}) become
\begin{equation}
\label{eq:massr}
\dot{M} = 4\pi r^2 \dot{m} 
= 4\pi r^2 (\rho_0 v_r + \left < \rho^{\prime} v^{\prime}_r\right > ) \, ,
\end{equation}
\begin{equation}
\label{eq:momentumr}
\dot{m} \, \partial_r v_r = 
-\partial_r P_0
+\rho_0 g_r 
- \partial_r \left < \rho R_{rr} \right > 
- \frac{2 \left <\rho R_{rr} \right > }{r} 
+ \frac{\left < \rho (R_{\theta \theta} + R_{\phi \phi}) \right >}{r}\, ,
\end{equation}
and 
\begin{equation}
\label{eq:entropyr}
\dot{m} \, \partial_r s_0 = 
\left < \frac{\rho \dot{q}}{T} \right >
+ \frac{\rho_0 \epsilon}{T_0}
- \nabla_r \cdot \left < F_r \right > \, ,
\end{equation}
where the $r$, $\theta$, and $\phi$ subscripts refer to the radial and
angular components in spherical coordinates.  Therefore, $\partial_r$
is the partial derivative with respect to $r$, and $\nabla_r \cdot$ is the radial
part of the divergence.  Since we assume steady state, the mass
accretion rate, $\dot{M}$, is a constant.  For isotropic turbulence, the last three
terms of eq.~(\ref{eq:momentumr}) reduce to $\partial_r \left <
  \frac{2}{3} \rho K \right > $, the gradient of turbulent pressure.  However, \citet{arnett09} note that
buoyancy-driven turbulence in spherical stars is not isotropic but is
most consistent with $R_{rr} = R_{\theta \theta} + R_{\phi \phi}$ and $R_{\theta \theta} =
R_{\phi \phi}$.  In this work, we adopt this later assumption where
convenient, but we retain the general expression in
eq.~(\ref{eq:momentumr}) as a reminder that the relationships among the
Reynolds stress components must be determined by theory,
simulation, or experiment.

The equivalent steady-state and spherically symmetric equations for Reynolds stress,
entropy flux, and entropy variation are 
\begin{equation}
\begin{array}{lll}
\lefteqn{v_r \partial_r \left < \rho R_{rr} \right > 
+ \left < \rho R_{rr} \right > \nabla_r \cdot v_r = } \\
& & + 2 \left < \rho^{\prime} v^{\prime}_r \right > g_r
- 2 \left < \rho R_{rr} \right >\partial_r v_r \\
& & - \left <  v^{\prime}_r \partial_r P^{\prime} + \partial_r P^{\prime}
v^{\prime}_r \right > 
- \nabla_r \cdot \left < \rho v_r^{\prime}R_{rr} \right > 
- \rho_0 \varepsilon_{rr} \, , \\
\end{array}
\end{equation}
\begin{equation}
v_r \partial_r \left < \rho R_{\theta \theta} \right > 
+ \left < \rho R_{\theta \theta} \right > \nabla_r \cdot v_r =
- \nabla_r \cdot \left < \rho v_r^{\prime}R_{\theta \theta} \right > 
- \rho_0 \varepsilon_{\theta \theta} \, ,
\end{equation}
\begin{equation}
v_r \partial_r \left < \rho R_{\phi \phi} \right > 
+ \left < \rho R_{\phi \phi} \right > \nabla_r \cdot v_r = 
- \nabla \cdot \left < \rho v_r^{\prime} R_{\phi \phi} \right > 
- \rho_0 \varepsilon_{\phi \phi} 
\end{equation}
\begin{equation}
\label{eq:fluxr}
\begin{array}{lllr}
\lefteqn{v_r \partial_r \left < F_r \right > + \left < F_r \right> \nabla_r
\cdot v_r = } \\
& & + \rho_0 \eta \left < Q \right > g_r \\
& & -\left < F_r \right > \partial_r v_r 
- \left < \rho R_{rr} \right > \partial_r s_0 \\
& & - \left < s^{\prime} \partial_r P^{\prime}\right > \\
& & - \partial_r \left < F_r v_r^{\prime}\right >
- 2 \left< F_r v_r^{\prime}\right > / r
+ \left < F_{\theta} v^{\prime}_{\theta} + F_{\phi}
    v^{\prime}_{\phi} \right > / r \\
& & + \left< \rho v_r^{\prime} \dot{q} / T \right > \, , \\
\end{array}
\end{equation}
and
\begin{equation}
\begin{array}{lllr}
\lefteqn{\rho_0 v_r \partial_r Q + \nabla_r \cdot \left ( \rho_0 v_r
  \right )
=} \\
& &
 -2 \left < F_r \right > \partial_r s_0 
+ 2 \rho_0 \eta \left < Q \right > \dot{q}/ T_0 
- \nabla_r \cdot \left < \rho v_r^{\prime} Q \right >\, . \\
\end{array}
\end{equation}
Finally, the spherically symmetric kinetic energy equation is
\begin{equation}
\begin{array}{llll}
\lefteqn{v_r \partial_r \left < \rho K \right > + \left < \rho K \right >
\nabla_r \cdot v_r = } 
\\
& & &
+ \left < \rho^{\prime} v^{\prime}_r \right > g_r
+ \left < \rho R_{rr}\right > \partial_r v_r
\\
& & &
- \nabla_r  \cdot \left < F_{K} \right >
- \nabla_r  \cdot \left < F_{P} \right >
+ \left < P^{\prime} \nabla \cdot {\mathbi{v}^{\prime}}\right >
- \rho_0 \epsilon \, . \\
\end{array}
\end{equation}

\section{Characterizing Turbulence of 2D Core-Collapse Simulations and
  Validating Averaged Equations}
\label{section:compare2d}

Having derived the mean-field equations and identified the important
turbulent correlations, we now characterize the background and turbulent
profiles of 2D simulations.  Most importantly, we validate that the
Reynolds averaged equations are consistent with the 2D results.
Section \ref{section:2dsimulations} briefly describes the 2D
simulations, highlighting general qualities that are relevant for turbulence
analysis such as the location and extent of turbulence and neutrino
heating and cooling.  Then in \S \ref{section:2dcorrelations}, we
characterize the turbulent correlations in the 2D simulations.
Finally, in \S \ref{section:validateequations}, we validate the
averaged equations. 
  
\subsection{2D Simulations}
\label{section:2dsimulations}

The 2D results presented here were calculated using BETHE-hydro
\citep{murphy08a} and are the same simulations that were used in \citet{murphy09} to develop a
gravitational wave emission model via turbulent plumes.  While
\citet{murphy09} considered a large suite of simulations, for clarity,
we focus on one simulation that simulated the collapse and explosion
of a solar metallicity, 15 \msun\ progenitor model \citep{woosley07} and used a driving neutrino luminosity of $3.7
\times 10^{52}$ erg s$^{-1}$.  See \citet{murphy08a} for more details
on the technique and \citet{murphy09} for the setup of this particular 2D simulation.

\begin{figure}[t]
\epssclone
\plotone{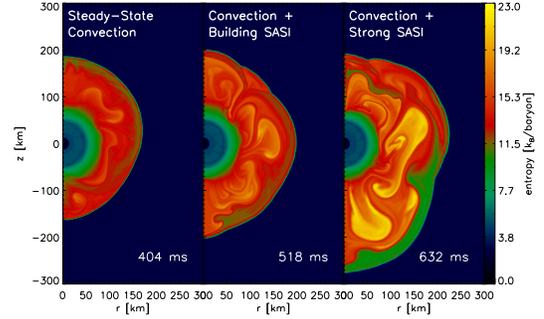}
\caption{Color map of entropy at times representing the three dominant
  phases.  The left panel shows the flow during steady-state
  convection (404 ms after bounce), the middle panel illustrates the dynamics and entropy
  distribution during the building convection and SASI stage (518 ms), and the
  far right panel shows the flow during strong convection and SASI
  (632 ms).  Most of the subsequent figures show specific turbulence
  characteristics at these three times.\label{convectionsasistills}}
\epsscale{1.0}
\end{figure}

To demonstrate the evolution of turbulence, most figures of this paper
highlight three phases after bounce.  These three stages correspond to
modest steady-state convection (404 ms), growing convection and SASI
(518 ms), and strong convection and SASI (632 ms), and the entropy color maps in
Fig. \ref{convectionsasistills} provide visual context for the shock
location, heating and cooling, and location and extent of the turbulence.

\begin{figure}[t]
\epsscltwo
\plotone{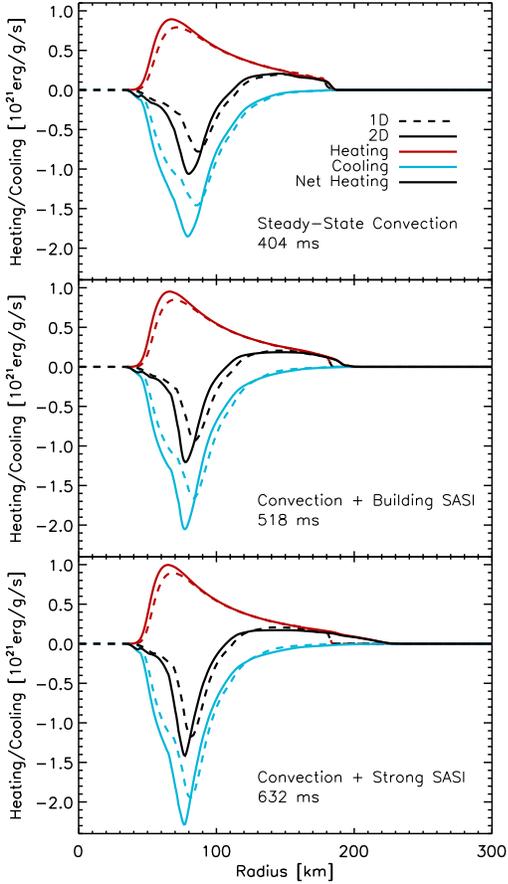}
\caption{Neutrino heating (red), cooling (blue), and net heating (black) profiles as a function of
  radius.  The times correspond to the stages of post-bounce evolution
  highlighted in Fig.~\ref{convectionsasistills}.  This plot compares the 1D
  (dashed lines) and 2D (solid lines) profiles.  Above the gain
  radius ($\sim$ 100 km), the net heating is positive.  In this gain
  region, a negative entropy gradient formally drives buoyant
  convection.  Below the gain region, a positive entropy gradient
  stabilizes against convection.   Though only the gain region is
  formally convective, overshoot carries convection well into the
  stable region below (up to
  $\sim$40 km below).  By definition, the
  heating profiles are similar.  The cooling profiles show some minor
  differences, but these lead to minor differences in the entropy profiles (see Fig. \ref{entropy3times}).  In
  Fig.~\ref{entropy3times}, we show that the most notable differences
  between 1D and 2D are produced by the divergence of the convective
  entropy flux.} \label{heating3times}
\epsscale{1.0}
\end{figure}

Our focus is on the most obvious turbulent region, which
extends from $\sim$80 km to the shock ($\gtrsim$180 km).  This
postshock turbulence is driven by neutrino heating, and in
Fig. \ref{heating3times}, we show neutrino heating (red), cooling
(blue), and net heating (heating minus cooling, black lines) profiles
for 1D (dashed-lines) and 2D (solid-lines) simulations.  These local
heating and cooling rates are calculated using eqs.~(4-5) of
\citet{murphy08b} and a neutrino luminosity of $L_{\nu_e} = 3.7 \times
10^{52}$ erg s$^{-1}$. Below the gain radius, $\sim$100 km, cooling
dominates heating, but above
the gain radius, heating dominates cooling.  This latter region is
called the gain region and drives turbulent convection.  After
matter accretes through the shock, it advects downward through the
gain region, producing a negative entropy gradient.  In turn, this
negative entropy gradient drives buoyant convection.  Though the region below the gain radius has
a positive entropy gradient and is formally stable to convection,
momentum carries plumes well into the cooling region \citep{murphy09}.
This is a well known phenomenon in stellar convection and is called
overshoot.  In neutrino-driven convection, the depth of overshoot can
be quite large, $\sim$20-40 km, \citep{murphy09}.

\subsection{Turbulent Correlations of 2D Simulations}
\label{section:2dcorrelations}

\begin{figure}[t]
\epsscltwo
\plotone{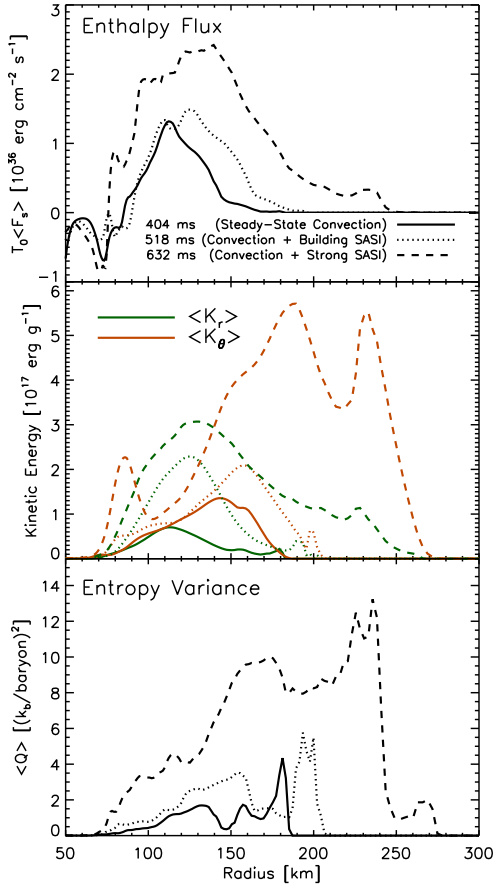}
\caption{Radial profiles for the most fundamental turbulent
  correlations: enthalpy flux, $T_0 \left < F_s \right > $, (top
  panel), kinetic energies, $K_r$ \& $K_{\theta}$, (middle panel), and
  entropy variance, , (bottom panel).  The times correspond to the
  three postbounce phases shown in Fig. \ref{convectionsasistills}.   These plots
show that turbulence grows with time, and the radial profiles
indicate that turbulence is dominated by non-local evolution of
coherent large-scale buoyant plumes (see text for more details).
A simple model that considers the evolution of sinking low entropy
plumes explains the features of $T_0 \left < F_s \right >$ and $\left
< K_r \right >$ and rising high entropy plumes explain $\left <
Q \right >$.  See \S \ref{section:2dcorrelations} for an
explanation.  \label{fskinetic3times}}
\epsscale{1.0}
\end{figure}

\begin{figure}[t]
\epsscltwo
\plotone{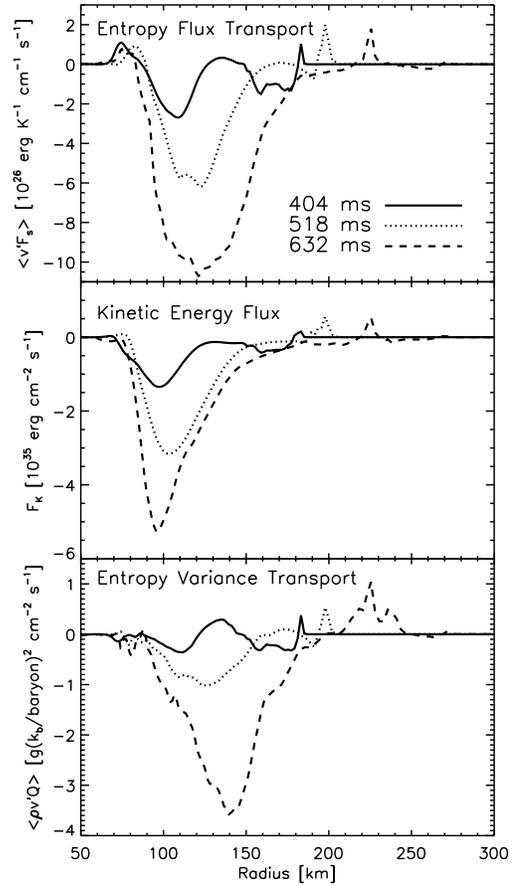}
\caption{Similar to Fig.~\ref{fskinetic3times} but for the transport
  of entropy flux, $\left < v^{\prime} F_s\right >$ (top
  panel), turbulent kinetic energy flux, $\left < F_K \right >$, and
  the entropy variance flux, $\left < \rho v^{\prime} Q\right >$.
  All transport terms are negative indicating that the transport of
  each correlation is dominated by the large speeds of the sinking
  plumes.  \label{transport3times}}
\epsscale{1.0}
\end{figure}

Figures \ref{fskinetic3times} and \ref{transport3times} show
the radial profiles of the primary correlations
in the averaged equations,
eqs.~(\ref{eq:mass}-\ref{eq:entropyvariance}).  The lowest moment
correlations are the turbulent enthalpy flux ($T_0 F_s$), the turbulent kinetic
energies ($K_r$ \& $K_{\theta}$; or Reynolds stresses), and the
entropy variance ($Q$), and all three are shown
in the top, middle, and bottom panels of Fig.~\ref{fskinetic3times}.
Other important higher order correlations are the turbulent transport
terms, which are the transport of entropy flux, $\left < v^{\prime}_r
F_s \right >$, the turbulent kinetic energy flux, $\left < F_K \right
>$, and the entropy variance flux, $\left < \rho v^{\prime}_r
Q \right > $.  The radial profiles of these higher order correlations
are in
Fig. \ref{transport3times}.  In general, all turbulent correlations
increase over time, and the radial profiles indicate that turbulence is dominated by coherent rising and sinking large-scale buoyant plumes.

The enthalpy (or entropy) flux has two broad physical interpretations.
Most obviously, the entropy flux indicates the direction
and magnitude of entropy transport due to turbulent motions.
Naturally, positive and negative entropy fluxes correspond to upward
and downward entropy transport.  In addition to indicating the
direction of entropy transport, the sign of the flux also indicates
the direction of the buoyancy forces driving the turbulence.
To understand this second interpretation, consider that the
entropy flux is defined using the correlation of the velocity and entropy perturbations (i.e. $\left < s^{\prime}
\mathbi{v}^{\prime}\right >$).  In regions where convection is
actively driven, high entropy plumes rise buoyantly and low
entropy plumes sink.  In other words, the entropy and velocity
perturbations are either both positive or negative.  Hence, the correlation and entropy flux are always
positive in regions of actively driven convection.  At boundaries,
where the plumes are decelerated due to a stable background, the
correlation or entropy flux, is always negative.  For example, as a
low entropy plume penetrates into the lower stable layer, the sinking
plume becomes immersed in a background that has even lower entropy.
While the velocity perturbation remains negative, the entropy
perturbation changes sign, becoming positive.  Consequently, the
correlation and entropy flux are negative in the bounding stabilizing
regions.

With these interpretations of the entropy (enthalpy) flux, the top panel of
Fig. \ref{fskinetic3times} shows where convection is
actively driven, where plumes are decelerated by bounding stable
regions, and the magnitude of each as a function of time.  At the
shock, the entropy flux is zero.  This is in contrast with the
results of \citet{yamasaki06}, who find an appreciable enthalpy flux
at the shock.  Whether the enthalpy flux is zero at the shock has
consequences for the stalled-accretion shock solution (See
\S~\ref{section:model1} for further discussion).  

Naively, one would expect the gain radius
($\sim$100~km) to mark the boundary between actively driven convection
in the heating region and the stabilizing effects in the cooling
layers below.  
Though this is roughly correct, careful inspection of the enthalpy profiles
show that the transition from positive to negative entropy flux does not correspond exactly with the gain
radius.  Instead, the gain radius corresponds best with the change in slope of
the enthalpy profile.  Above the gain radius, where convection is
actively driven, the enthalpy profile has a negative gradient, and
below the gain radius the gradient is positive.  

This
profile can be best understood considering a single low entropy plume
originating at the shock.
First of all, as this plume accelerates
downward, both the entropy and velocity perturbations grow in magnitude.  This explains
the negative entropy flux gradient above the gain radius.  Below the gain
radius the background entropy gradient is positive.  Therefore, the
entropy perturbation diminishes as the background entropy reduces to
the level of the low entropy plume.  At this point, both the entropy
perturbation and entropy flux are zero.  Consequently, enthalpy flux
gradient is positive below the gain radius.  With a negative gradient
above the gain region and a positive gradient below it, the enthalpy
flux is maximum at the gain radius.  As the plume's
inertia carries it beyond this radius, the enthalpy flux becomes negative in the
stabilizing layers.  These general characteristics are observed at all
times, but with increasing magnitude at later times when convection is
more vigorous.

The same simple model can explain the radial ($\left < K_r \right
>$, green lines) and tangential ($\left < K_{\theta} \right >$, orange
lines) kinetic energy profiles in the middle
panel of Fig. \ref{fskinetic3times}.  In fact, like the enthalpy flux,
$\left < K_r \right > $ peaks at the gain radius.  Because the sinking
plumes have higher speeds than the rising plumes and the kinetic
energy is weighted by the square of the speed, $\left < K_r \right >$
is dominated by the sinking plumes.  Again, the radial profile of
$\left < \right > $ is consistent with low entropy plumes originating
at the shock, accelerating downward to a maximum speed at the gain radius, and
decelerating in the stabilizing region below the gain radius.  This is also consistent with the results of
\citet{murphy09}.  The tangential component, $\left <
K_{\theta} \right >$, on the other
hand, shows a maximum just 10s of km away from the shock.  This is
where the rising plumes encounter material that has just passed
through the shock and both turn their trajectories in the $\theta$-direction.  

Interpretation of the entropy variance ($Q$ and the bottom panel of
Fig. \ref{fskinetic3times}) is less clear.  Like the enthalpy flux and
kinetic energies, $Q$ increases with time.  The most naive
interpretation is that the variation of total heating among the
sinking and rising plumes increases over time.  However, it is not obvious whether this increase in variance is due to more heating or less
cooling in either the rising or sinking plumes...or both.  While it
seems clear that the profiles of $T_0 F_s$ and $K_r$ are dominated by
the sinking plumes, there is circumstantial evidence that $Q$ might be
dominated by the behavior of rising plumes.  For one, there is a
gradual rise in $Q$ from the lower convective boundary to the upper
boundary, suggesting growth with the rise of buoyant plumes.  But the
most telling evidence comes from the entropy color maps of
Fig.~\ref{convectionsasistills}.  In these maps, it is obvious that
the entropy of the sinking plumes is roughly constant over time, while
the entropy of rising plumes increases with time.  Hence, the variance
of entropy increases with time because the maximum entropy of the
rising plumes increases.

All three transport terms in Fig.~\ref{transport3times}, $\left <
v^{\prime}_r F_s \right >$, $\left < F_K \right >$, and $\left < \rho
v^{\prime}_r Q \right >$, are negative nearly everywhere.  Hence, the
flow of core-collapse turbulence acts to transport entropy flux,
kinetic energy, and entropy variance downward.  This is typical of buoyancy driven convection and is observed in most
simulations of convection within stellar interiors \citep{cattaneo1991,meakin10}.
For the entropy flux,
and turbulent kinetic energy, this fact further
supports the notion that the turbulent correlations are dominated by
sinking plumes.  On the other hand, at first glance, the negative
transport of $Q$ seems to be at odds with our previous conclusion
that rising plumes dominate the character of $Q$.  However, the moment
of the entropy variance, $\left < Q
\right >$, weights only the variance in the entropy, but the entropy
variance flux weights the velocity of the plumes as well.  In general,
the speed of the sinking plumes is larger than the speed of rising
plumes.  Consequently, while rising plumes provide the most weight to
$\left < Q \right >$, sinking plumes provide the greatest weight to
$\left < \rho v^{\prime}_r Q \right >$.  In \S \ref{section:comparemodels}, we present
simple models for these transport terms that assume the dominance of
sinking plumes.

\subsection{Comparing Averaged Background Equations with 2D Simulations}
\label{section:validateequations}

Having presented the turbulent correlations of 2D core-collapse
simulations, we now validate the Reynolds-averaged equations.
Specifically, we validate the spherically symmetric background
equations, eqs.~(\ref{eq:massr}-\ref{eq:entropyr}).  For the sake of
brevity, we do not show a plot of mass conservation but simply report
that $\dot{M} = 4 \pi r^2 (\rho_0 v_r + \langle \rho^{\prime}
v^{\prime}_r \rangle)$ is indeed satisfied in the 2D simulations.

\begin{figure}[t]
\epsscltwo
\plotone{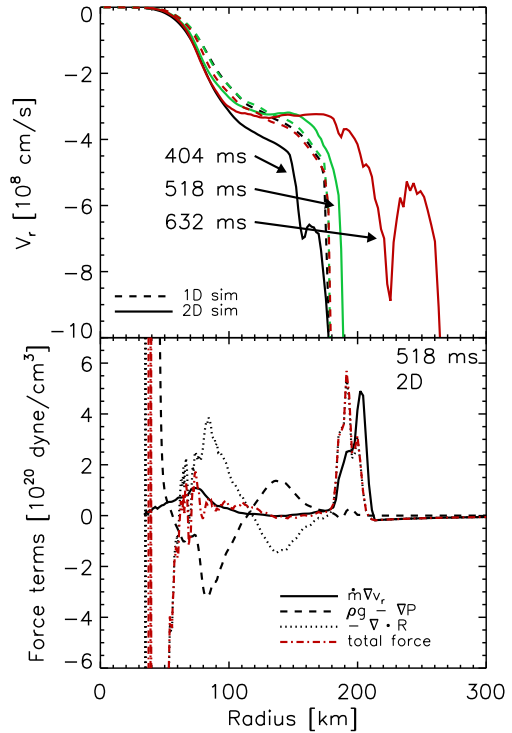}
\caption{Radial velocity (top panel) and force (bottom panel) profiles
  for the three phases shown in Fig. \ref{convectionsasistills}.  The
  forces are the terms in the momentum equation,
  eq.~(\ref{eq:momentumr})  TOP PANEL:  The 1D velocity profiles for the three
  phases are quite similar.  The 2D velocity profiles evolve as the
  shock radius and convective vigor increase.  At later times when
  convection is strongest, the velocity gradient tends to zero in the
  convective region. BOTTOM PANEL:  We compare the individual
  force terms of the momentum equation,  eq.~(\ref{eq:momentumr}).
  The excess force between gravity and the pressure gradient (dashed)
  is nearly balanced by the divergence of the Reynolds stress
  (dotted). The red dashed line shows that the sum of these forces do
  indeed equal the actual force, $\dot{m} \nabla v_r$ (solid curve).
  This validates the form of the momentum equation,
  eq.~(\ref{eq:momentumr}), including the convective
  term. \label{velocity3times}}
\epsscale{1.0}
\end{figure}

Figure \ref{velocity3times} validates the form of the momentum equation,
eq. (\ref{eq:momentumr}), including the turbulence terms.  In the top
panel, we plot the velocity profile of 1D and 2D simulations as a function of radius, and in the bottom panel, we plot the
dominant force terms in the momentum equation for the 2D simulations only.
Specifically, the bottom panel shows the difference in the
gravitational and pressure gradient forces, $\rho_0 \mathbi{g} -  \nabla P_0$ (dashed-line), and the
divergence of the Reynolds stress, $\nabla \cdot \mathbf{R}$ (dotted line).
The solid black line shows $\dot{m} \nabla_r v_r$ from the 2D
simulation.  This last term is the left-hand side of
eq. (\ref{eq:momentumr}), and in steady state, represents the total
force per unit area that a Lagrangian parcel of matter experiences. 
If eq.~(\ref{eq:momentumr}) represents the correct derivation of the
momentum equation including turbulence terms, then the sum of the
right-hand side terms (dot-dashed red line) should equal the solid
black line.  Away from the shock, they agree quite well.
Interestingly, the right-hand side
is essentially zero in the heating region where convection is actively
driven.  This implies that the difference in the gravitational force
and the pressure gradient is nearly balanced by the divergence of the
Reynolds stress. 

\begin{figure}[t]
\epsscltwo
\plotone{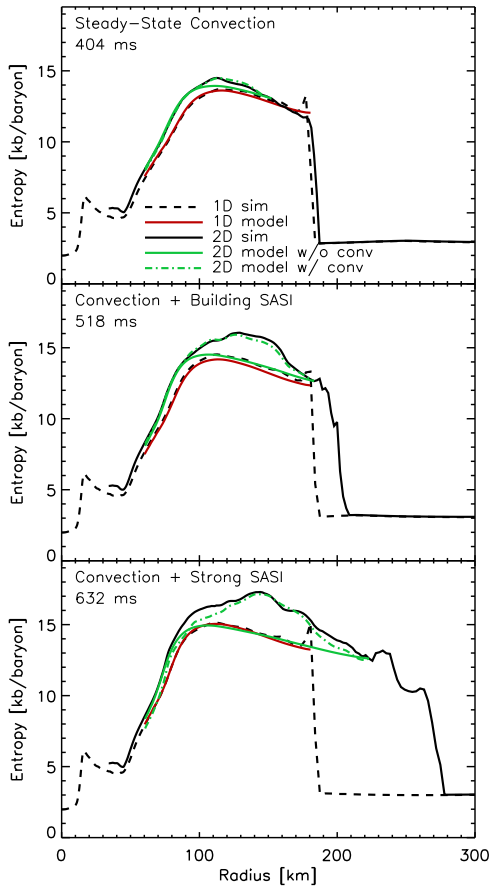}
\caption{Radial entropy profiles during the three phases shown in Fig
  \ref{convectionsasistills}.  In these plots, we compare the results
  from 1D (dashed, black curves) and 2D (solid, black curves)
  simulations to integrations of the entropy equation,
  eq.~(\ref{eq:entropyr}). The
  red curve shows the result of this integration using background
  profiles from the 1D simulation and no convective terms.  The
  agreement with 1D simulations validates this technique to verify the
  equations.  The green curves show the results of integrating the
  entropy equation using background profiles from the 2D simulations
  without convective terms (solid green) and with convective terms
  (dashed green) curve.  Excluding the convective terms under-predicts the entropy profile,
  and including the convective terms raises the entropy and produces accurate entropy profiles.\label{entropy3times}}
\epsscale{1.0}
\end{figure}

Figure \ref{entropy3times} validates the Reynolds-averaged entropy
equation, eq.~(\ref{eq:entropyr}).  The solid black line corresponds to the 2D
results, and the black dashed line shows the results of 1D
simulations.  For comparison, the red curve is the integration of
eq.~(\ref{eq:entropyr}) using the 1D density, velocity, and heating
profiles.  Since the 1D simulations are not able to simulate multi-dimensional effects, we omit the turbulent dissipation and the
entropy flux terms in this integration.  The remarkable agreement
between the results of this integration and the 1D simulation
bolsters our approach in validating eq.~(\ref{eq:entropyr}).  Similarly,
the solid green line shows the integration of eq.~(\ref{eq:entropyr})
using the 2D density, velocity, and heating curves and no convection
terms.  This curve is similar to the 1D results and clearly under
predicts the entropy in the gain region.  On the other hand, including
the turbulence terms in the integration of eq.~(\ref{eq:entropyr})
(dot-dashed green curve) dramatically improves the comparison.
Therefore, given the right entropy flux and turbulent dissipation, we
conclude that eq.~(\ref{eq:entropyr}) accurately determines the background flow.
Additionally, we conclude that the
 turbulence models of \S \ref{section:models} must, at a minimum,
 produce accurate entropy flux profiles to accurately describe the
effects of convection in 2D simulations.

\begin{figure}[t]
\epssclthree
\plotone{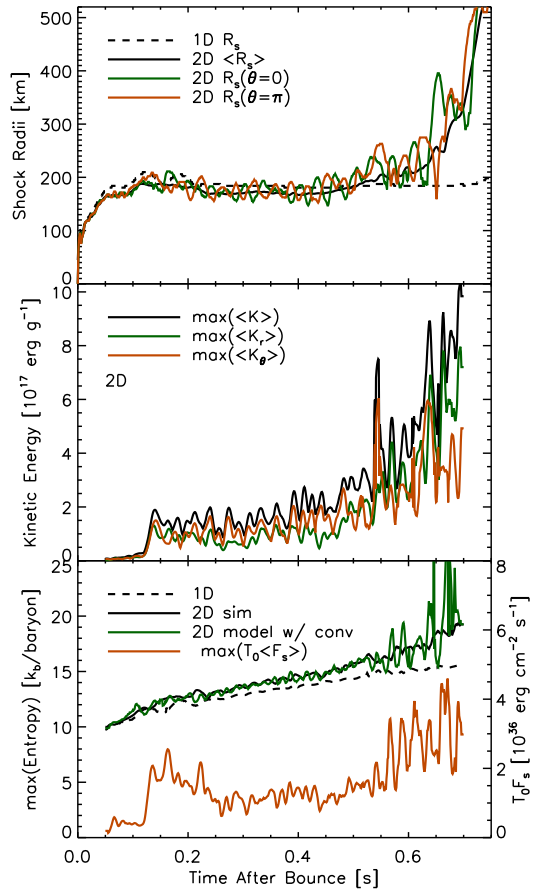}
\caption{A comparison of shock radii and convection diagnostics as a
  function of time after bounce.  TOP PANEL:  We plot shock
  radii as a function of time after bounce.  We show the 1D and average 2D shock radii using
  dashed and solid black lines, respectively.  Before the onset of
  vigorous SASI and convection ($\sim$550 ms), both stall at $\sim$180
  km.  Afterward, the 2D average shock radius begins to
  climb to $\sim$320 km, at which point all measures of shock radii
  expand outward in an explosion.
  To illustrate the asymmetries in the shock we also plot the shock
  radius at the poles ($\theta$=0 and $\theta$ = $\pi$).  The shock
  radii at the poles oscillate about the average shock position until
  explosion ($\sim$700 ms).  MIDDLE PANEL: The middle panel plots the maximum of
  the total, radial, and transverse kinetic energies. BOTTOM PANEL: The bottom
  panel shows the maximum entropy of the 1D and 2D simulations, the
  maximum entropy resulting from integrating eq.~(\ref{eq:entropyr}),
    and the maximum enthalpy flux $T_0\left < F_s \right >$. Figure~\ref{fskinetic3times} samples these convection
  diagnostics at three times and indicates that convection grows
  monotonically.  These plots confirm that convection indeed grows
  with time.  Furthermore, Fig. \ref{entropy3times} validates
  the Reynolds-averaged entropy equation, eq.~(\ref{eq:entropyr}), at
  three specific times, and the bottom panel validates this equation at
  all times.
 \label{convectiondata}}
\epsscale{1.0}
\end{figure}

Figures \ref{fskinetic3times}-\ref{entropy3times} suggest that convection grows monotonically,
but these figures sparsely sample convection and its effects at three
times.  In Fig. \ref{convectiondata}, we illustrate that
convection indeed grows monotonically from core bounce until explosion.  
To provide some context with shock position, we plot in the top panel shock radii as a function of
  time after bounce.  We show the 1D and average 2D shock radii using
  dashed and solid black lines, respectively.  Before the onset of
  vigorous SASI and convection ($\sim$550 ms), both stall at $\sim$180
  km.  Afterward, the 2D average shock radius climbs to $\sim$320 km,
  at which point all measures of the shock unambiguously expand in an explosion.
  To illustrate the asymmetries in the shock we also plot the shock
  radius at the poles ($\theta$=0 and $\theta$ = $\pi$).  The shock
  radii at the poles oscillate about the average shock position until
  explosion ($\sim$700 ms).  The middle panel shows the maximum of
  the total, radial, and transverse kinetic energies.  In general, the
  turbulent kinetic energy steadily grows until explosion.  Finally, the
  bottom panel shows the maximum of 1D (black) and 2D (green) average
  entropy profiles.  For comparison, we plot the maximum enthalpy flux
  in orange and the maximum entropy resulting from
  integrating eq.~(\ref{eq:entropyr}) in green. As in Fig.~\ref{entropy3times}, the 2D
  simulation consistently shows higher entropy at all times, and the
  results of integrating eq.~(\ref{eq:entropyr}) are consistent with the simulations.

In conclusion, the results of Figs.~(\ref{fskinetic3times}-\ref{convectiondata}) indicate that Reynolds
decomposition of the hydrodynamics equations, eqs. (\ref{eq:mass}-\ref{eq:entropy}), is
consistent with 2D simulations.  

\section{Turbulence Models}
\label{section:models}

In \S~\ref{section:reynoldsdecomp}, we derived the turbulence
equations, eqs.~(\ref{eq:mass}-\ref{eq:entropyvariance}) and showed that they suffer from a closure problem.
Therefore, finding solutions to the background and turbulence
equations requires a turbulence closure model.  In this section, we
present several turbulence models and identify their strengths and weaknesses.

In sections \ref{section:model1}-\ref{section:model4},
we present four turbulence models.  These models, with the exception
of Model 1, require a model for turbulent dissipation and transport.  Therefore, our
discussion of the models begins in
sections~\ref{section:dissipationmodel} \&
\ref{section:transportmodels} with
a model for turbulence dissipation and transport.  The next four sections, \S
\ref{section:model1}-\ref{section:algebraic}, present models for the
primary 2$^{\rm nd}$ order turbulent correlations, $F_s$, $K$, and
$Q$.  The first model, Model 1, is a reproduction of a model presented
by \citet{yamasaki06} and assumes that both $\left < \dot{s} \right >$
and $\nabla s_0$ are zero.  The resulting model is simple, but it
provides no equation for the turbulent kinetic energy.  Model 2 is a
closure model for the Reynolds stress, entropy flux, and entropy
variance and has been designed and calibrated to simulate isolated
buoyant plumes.  Model 3 is an algebraic model which is akin to MLT.

To model the higher order correlations, Models 2 \& 3 use expressions
involving the local values of 2$^{\rm{nd}}$-order correlations.  Hence,
some of the most important terms in a nonlocal problem are modeled
using local approximations.  While these local models are adequate in
some locations, they can be factors off in other locations.  Rather than relying
on these local models, we develop in Model 4 (\S~\ref{section:model4})
a novel global model that uses global conservation laws to constrain
the scale of convection.

Later, in \S~\ref{section:comparemodels}, we compare the results of
these models to the turbulent kinetic energies and entropy flux of the
2D simulations.

\subsection{Turbulent Dissipation: Kolmogorov's Hypothesis}
\label{section:dissipationmodel}

Starting with Kolmogorov's hypotheses for turbulent dissipation,
\citet{arnett09} construct a model for turbulent
dissipation which they validate with 2D and 3D stellar evolution calculations.
Buoyed by these successes, we construct a
similar model for turbulent dissipation.  One of the primary
hypotheses of Kolmogorov's theory is that turbulent energy is injected at the
largest scales and cascades to smaller scales.  Consequently, the rate
of turbulent energy dissipation is governed by the largest scales and
dimensionally is proportional to ${v^{\prime}}^3/\mathcal{L}$, where
$\mathcal{L}$ is an appropriate length scale, usually the
largest eddy size.  Therefore, our model for turbulent dissipation is
\begin{equation}
\label{eq:dissipation}
\epsilon = \frac{R^{3/2}}{\mathcal{L}} \, ,
\end{equation}
where $R = \mbox{tr}(\mathbf{R})$ is the trace of the Reynolds stress
tensor.  

In most astrophysical calculations, the length scale is assumed to be
proportional to the pressure scale height, $H_P = - (\partial \ln P /
\partial r)^{-1}$.  On the other hand, \citet{arnett09} found
that stellar convection fills the total available space and that this
also corresponds to the largest eddy size.  For very large convection
zones, they found that $\mathcal{L}$ is at most $4 H_P$.  Therefore, we set the
length scale as $\mathcal{L} = \mbox{max}(4 H_P,\mathcal{L}_{\mbox{conv}})$, where
$\mathcal{L}_{\mbox{conv}}$ is the size of the region unstable to convection.

Given Kolmogorov's hypothesis, this model for $\epsilon$ is the most
basic model that one can assume.   Later in \S~\ref{section:model4},
we propose a model for $\epsilon$ that satisfies Kolmogorov's
hypothesis but apparently represents the dissipation of buoyant
plumes via entrainment.  To satisfy global and local constraints for convection, we
find that $\epsilon$ is best modeled by a linear function of
distance from the shock.  In 3D simulations of stellar convection in
which negatively buoyant plumes dominate, \citet{meakin10} found a
similar result.  These results suggest that entrainment between rising
and sinking
plumes govern dissipation.

\subsection{Turbulent Flux Models}
\label{section:transportmodels}

\par The gradient diffusion approximation\footnote{This approximation
  is also referred to as the down gradient approximation, and refers
  to the closure whereby the flux of a quantity $F_i$ is proportional
  to the gradient of the quantities density $E_i$, through  $F_i
  \propto - \nabla E_i$.} has found wide spread usage for closing the
turbulent transport terms \citep[e.g.,][]{pope00,launder02}.  For
example, Model 2, which we introduce in
\S \ref{section:model2}, uses this assumption.  The theoretical basis
for this closure model is two fold: first, the transported
quantity behaves like a scalar; second, scale separation is a
good approximation in the sense that transport is mediated by
fluctuations on scales small compared to the largest scales
characterizing the turbulent flow \citep{dalyharlow1970}.  The
turbulent transport in a buoyant convection zone, however, is
generally recognized to be mediated by large scale coherent plumes
\citep[e.g.][and \S\ref{section:2dcorrelations}]{cattaneo1991}, thus
challenging the theoretical underpinning for a gradient diffusion
approximation.  Furthermore, the shortcomings of the gradient
diffusion approximation for thermal convection were explicitly
illustrated through a series of 3D simulations of turbulent stellar
interiors by \citet{meakin10}.

\par Rather than a gradient diffusion approximation, we propose flux
models that are proportional to  ${R_{rr}}^{1/2} E_i$, where $E_i$ is
the density of the transported quantity. This model is built on the
advective nature of the transport in which $F_i \propto v' E_i$.  We
propose the following models for turbulent transport of the entropy
flux, turbulent kinetic energy, and entropy variance 
\begin{equation}
\label{eq:entropyfluxtransportmodel}
\left < v^{\prime}_r F_s\right > \approx - R_{rr}^{1/2} F_s,
\end{equation}
\begin{equation}
\label{eq:fkmodel}
\left < F_K \right > \approx - \rho_0 R_{rr}^{3/2},
\end{equation}
\noindent and
\begin{equation}
\label{eq:entropyvariancetransportmodel}
\left < \rho v^{\prime}_r Q \right > \approx - \frac{1}{2}\rho_0 R_{rr}^{1/2} Q.
\end{equation}
Except for the entropy variance flux, the comparison with 2D
simulations (\S \ref{section:comparemodels} and
Fig. \ref{transportmodels}) indicates that the constant of
proportionality is $\sim$1. The constant of proportionality for the
entropy variance flux is found to be $\sim$1/2.

\subsection{Model 1: Steady-State and Zero Entropy Gradient Model}
\label{section:model1}

The first turbulence model that we consider is a simple model for $F_s$
presented by \citet{yamasaki06}.  \citet{yamasaki06} assumed steady
state and that the entropy gradient in eq.~(\ref{eq:entropyr}) is
zero.  From these assumptions, they derived a simple differential
equation for the enthalpy flux.  Here, we reproduce this equation, expressing it in terms of the entropy flux:
\begin{equation}
\label{eq:forfconv}
\nabla_r \cdot F_s = \frac{\dot{Q}}{T_0} \, .
\end{equation}
A simple integral of this equation with a boundary condition leads to
an expression for the entropy flux.  \citet{yamasaki06} assumed zero
flux at the lower boundary and integrated upward resulting in a
nonzero flux at the shock.

A major advantage of this model is that it is simple and
straightforward to find a solution for $F_s$.  
Unfortunately, the assumptions which lead to
  simplicity also lead to inaccurate turbulent profiles (see \S~\ref{section:comparemodels}).  For one, there is no solution
for the kinetic energies.  Secondly, this model completely ignores the
velocity of the background flow.  As we show in \S~\ref{section:comparemodels} and Fig.~\ref{compsimmodels_fs}, this is inconsistent with the characteristics of neutrino-driven convection in 2D simulations.
Thirdly, this model produces flawed entropy flux profiles.  For
example, it results in a nonzero entropy flux at the shock, while 2D simulations show zero entropy flux
at the shock (Figure~\ref{fskinetic3times}). 
In fact, the entropy flux is zero at both
boundaries.  To compensate for this nonzero flux at the shock,
\citet{yamasaki06} modified the shock jump conditions.  However, the
solutions to eqs.~\ref{eq:mass}-\ref{eq:entropyvariance} are quite sensitive to the form of the boundary
conditions at the shock.  Incorrect boundary conditions will lead to
erroneous solutions.  

\subsection{Model 2: Reynolds Stress and Heat Flux Closure Model}
\label{section:model2}

Model 2 is a Reynolds stress and heat flux closure model that has been designed and calibrated to model isolated
buoyant plumes \citep{launder02}.  Though core-collapse turbulence
involves more than just isolated plumes, neutrino-driven
convection is in fact buoyancy-driven.  Therefore, it is plausible that the
Reynolds stress and heat flux model might be an appropriate closure model.  

The model for the Reynolds stress equation, eq.~(\ref{eq:reynoldsstress}), is
\begin{equation}
\label{eq:reynoldsstressmodel}
\nabla \cdot (\rho \mathbi{v} \otimes \mathbf{R}) = \rho (
\mathbf{G} + \mathbf{P} + \boldsymbol{\Pi} + \mathbf{D} - \frac{2}{3}
\delta_{ij} \epsilon) \, ,
\end{equation}
where $\mathbf{G}$ and $\mathbf{P}$ are the buoyant and shear
production terms, $\boldsymbol{\Pi}$
is the pressure-strain term, $\mathbf{D}$ is the diffusion
term, and the final term is turbulent dissipation.  All but the first
two terms require models, and the expressions for the first two
production terms are easily read from eq.~(\ref{eq:reynoldsstress}).

The first of the modeled terms is the pressure-strain correlation, and
it acts to redistribute energy among the Reynolds stress components.
For buoyant flows, the pressure-strain correlation is generally modeled
by three contributions
\begin{equation}
\boldsymbol{\Pi} = \boldsymbol{\Pi}_1 + \boldsymbol{\Pi}_2 + \boldsymbol{\Pi}_3 \, ,
\end{equation}
where the first term is proportional turbulent stress, the
second term is proportional to the interaction between turbulent stress
and the mean strain, and the last term is proportional to buoyancy.
Explicitly, they are
\begin{equation}
\boldsymbol{\Pi}_1 = -c_1 \frac{\epsilon}{K} \left ( \mathbf{R} - \frac{2}{3}
\delta_{ij} \rho K \right ) \, ,
\end{equation}
\begin{equation}
\boldsymbol{\Pi}_2 = - c_2 \left ( \mathbf{P} - \frac{1}{3} \delta_{ij}
\mbox{tr}(\mathbf{P}) \right ) \, ,
\end{equation}
and
\begin{equation}
\boldsymbol{\Pi}_3 = -c_3 \left ( \mathbf{G} - \frac{1}{3} \delta_{ij}
\mbox{tr}(\mathbf{G}) \right ) \, ,
\end{equation}
where the parameters $c_1$, $c_2$, and $c_3$ are 3, 0.3, and 0.3 respectively.

\citet{launder02} suggest using the gradient-diffusion hypothesis to
model the diffusion term.  Specifically,
\begin{equation}
\mathbf{D} = c_R \nabla \cdot \left ( 
\frac{K}{\epsilon} \mathbf{R} \cdot \nabla \mathbf{R}
\right ) \, ,
\end{equation}
where the constant $c_R$ has been calibrated by experiments and
simulations to be 0.22.  This diffusion term inherently assumes that
the Reynolds stress (or kinetic energy) flux is proportional to the
gradient in the Reynolds stress.  This is most likely relevant in
shear dominated flows.  However, the buoyancy dominated flows of
core-collapse convection are characterized by large scale plumes.  As
a consequence, the transport of kinetic energy flux is directly proportional to
these bulk turbulent motions (i.e. $R^{3/2}$) rather than the gradient.

The final term to be modeled is the turbulent dissipation, $\epsilon$.
\citet{launder02} add another differential equation that includes more
production, diffusion, dissipation terms, and constants to be
calibrated.  To avoid these complications, we simply adopt the
turbulence model of \S~\ref{section:dissipationmodel}.

To derive the model turbulent kinetic energy equation, we take the
trace of the Reynolds stress model equation,
eq.~(\ref{eq:reynoldsstressmodel}),  
\begin{equation}
\label{eq:kineticmodel}
\begin{array}{llll}
\lefteqn{\mathbi{v} \cdot \nabla \left < \rho K \right > + \left < \rho K
\right > \nabla \cdot \mathbi{v} = } 
\\
& & & 
+ \left < \rho^{\prime} \mathbi{v}^{\prime} \right > \cdot \mathbi{g}
- \mbox{tr}\left ( \left < \rho \mathbf{R}\right > \cdot \nabla \mathbi{v}\right )
\\
& & & 
c_R \nabla \cdot \left ( 
\rho \frac{K}{\epsilon} \mathbf{R} \cdot \nabla K
\right )
- \rho_0 \epsilon \, ,
\\
\end{array}
\end{equation}
Except for two differences, this model kinetic energy equation is very similar to the full kinetic
energy equation, eq.~(\ref{eq:kinetic}).  The first difference is that
the work done by turbulent pressure perturbations, $\left < P^{\prime} \nabla \cdot \mathbi{v}^{\prime}  \right >$, is
absent in the model equation.  Because the pressure-strain correlation, $\boldsymbol{\Pi}$, is designed to only redistribute energy among the
components, the trace of this term is identically zero.  Hence, the
model assumes that this term is zero.  Because low Mach number
turbulence typically has negligible pressure perturbations, this is a
standard assumption in turbulence modeling.  Our 2D simulations
confirm this assumption.  The second difference is that $F_K$ is
assumed to be proportional to $\nabla K$.  This is a consequence of
the gradient-diffusion approximation, but in
\S~\ref{section:comparemodels}, we find that $F_K$ is not proportional
to the gradient but is best modeled as $F_K \propto R^{3/2}$.

The model for the entropy flux equation, eq.~(\ref{eq:entropyflux}), is
\begin{equation}
\label{eq:entropyfluxmodel}
\begin{array}{llll}
\lefteqn{
\mathbi{v} \cdot \nabla \mathbi{F}_s 
+ \mathbi{F}_s (\nabla \cdot \mathbi{v}) =
}
\\
& & & \rho_0 \eta Q \mathbi{g}
- \mathbi{F}_s \cdot \nabla \mathbi{v} 
- \mathbf{R} \cdot \nabla s_0
\\
& & & + c_s \nabla \cdot \left ( \frac{K}{\epsilon} \mathbf{R} \cdot \nabla
\mathbi{F}_s \right ) 
+ \boldsymbol{\Pi}_s
\\
\end{array}
\end{equation}
where $c_s = 0.15$ and the pressure-entropy correlation term
($\boldsymbol{\Pi}_s$) is the analog of the pressure-strain correlation
and is modeled by three terms 
\begin{equation}
\boldsymbol{\Pi}_s = \boldsymbol{\Pi}_{s1} + \boldsymbol{\Pi}_{s2} +
\boldsymbol{\Pi}_{s3} \, .
\end{equation}
In general, these terms act to dissipate $\mathbi{F}_s$ and represent pure turbulence,
turbulence and mean strain, and buoyancy interactions:
\begin{equation}
\boldsymbol{\Pi}_{s1} = -c_{1s} \frac{\epsilon}{k} \mathbi{F}_s \, ,
\end{equation}
\begin{equation}
\boldsymbol{\Pi}_{s2} = c_{2s} \mathbi{F}_s \cdot \nabla \mathbi{v} \, ,
\end{equation}
and
\begin{equation}
\boldsymbol{\Pi}_{s3} = c_{3s} \rho_0 \eta Q \mathbi{g} \, ,
\end{equation}
where the constants are $\{c_1s, c_2s, c_3s\} = \{ 2.85, 0.55,
0.55\}$.  
Once again, the transport term in eq.~(\ref{eq:entropyfluxmodel}) is
modeled using the gradient-diffusion approximation.

The final equation models the equation for entropy
variance, eq.~(\ref{eq:entropyvariance}):
\begin{equation}
\label{eq:entropyvariancemodel}
\rho \mathbi{v} \cdot \nabla Q + Q \nabla \cdot (\rho \mathbi{v}) = -2
\mathbi{F}_s \cdot \nabla s_0 
c_Q \nabla \cdot \left ( 
\rho \frac{K}{\epsilon} \mathbf{R} \cdot \nabla Q
\right )
- \frac{Q \rho \epsilon}{K r}
\end{equation}
where $c_Q = 0.11$ and $r = 0.56$.

Equations (\ref{eq:reynoldsstressmodel}-\ref{eq:entropyvariancemodel})
represent a model for the turbulence equations that is complete and
has been extensively tested and calibrated with experiment and
simulations \citep{launder02}.  Unfortunately, there are several
disadvantages to using this turbulence model.
For one, this model depends upon a large number of calibrated
constants.  In addition, these equations were designed
and calibrated to calculate the profile of a single isolated buoyant
plume.  The macroscopic flow of fully developed convection is very different.  Therefore, it is quite possible that the closure
relations presented in this model are not appropriate for core-collapse
convection.  Furthermore,
the dissipation terms make
eqs.~(\ref{eq:reynoldsstressmodel}-\ref{eq:entropyvariancemodel}) a
stiff numerical problem, requiring careful numerical
treatment.  Consequently, the solutions to these equations are extremely
sensitive to the uncertain dissipation models.

\subsection{Model 3: An Algebraic Model}
\label{section:algebraic}

In general, there are two strategies to finding turbulent solutions
(In \S~\ref{section:model4}, we present a third).
In the first, the turbulent correlations are solutions of differential
equations.  Models 1 \& 2 are of this type.  Within the second strategy,
a few key assumptions allow the differential
equations to be converted into a set of algebraic equations, and the turbulence correlations are solutions to these
algebraic equations.  An example of a turbulence model which uses a
set of algebraic equations is the mixing length theory (MLT); it is
used extensively through out astrophysics, including stellar structure
calculations \citep{kippenhahn90}.  

To transform the differential equations into algebraic equations, the
temporal and spatial derivatives must either vanish or be approximated
by algebraic expressions.  As an example of the first method, one can
assume that the temporal and advective derivatives are zero, (the
L.~H.~S. of the evolution equations
eqs. (\ref{eq:reynoldsstress}-\ref{eq:entropyvariance})).
This assumption is valid in most circumstances because even though the
terms on R.H.S. are large they often sum to zero.  Therefore, the
primary assumptions of this approach are steady state and local
balancing.  In the second approach, the spatial derivatives are
replaced with ratios of the variable to be differentiated and a length scale.
For example, the divergence of the kinetic energy flux
is roughly, $\nabla \cdot F_K \sim F_K / L$.  In effect, a global
boundary value problem is reduced to a set of local algebraic
expressions.

In some respects, these algebraic equations still retain nonlocal
characteristics.  The nonlocality is merely
hidden in the assumptions.  For example, in MLT, local gradients are
used to calculate the local buoyancy force, but the eddies are assumed
to remain coherent until a mixing length at which point they dissipate
their energy.  Hence, the finite size of the mixing length is an echo
of the true nonlocality of turbulence in a local prescription.  Where
local balancing is important, such as the heating region, these
approximations can give reasonable solutions.  However, in regions
where nonlocal transport is important, such as overshoot regions,
these algebraic models fail completely.

To derive an algebraic model, we assume that transport balances
buoyant driving.  Applying this assumption to the exact 2$^{\rm{nd}}$ moment
equations, eqs.~(\ref{eq:kinetic}-\ref{eq:entropyvariance}), and using
the flux models,
eqs.~(\ref{eq:entropyfluxtransportmodel}-\ref{eq:entropyvariancetransportmodel}),
of \S \ref{section:transportmodels} results in approximate kinetic
energy,
\begin{equation}
\label{eq:transporteqdriving1}
- \nabla_r \cdot \left ( \rho_0 R^{3/2}_{rr} \right ) \sim \eta F_s g
\, ,
\end{equation}
entropy flux
\begin{equation}
\label{eq:transporteqdriving2}
- \nabla_r \cdot \left ( R^{1/2}_{rr} F_s \right ) \sim \frac{1}{2}
\rho_0 \eta g Q \, ,
\end{equation}
and entropy variance equations,
\begin{equation}
\label{eq:transporteqdriving3}
\nabla_r \cdot \left ( \rho_0 R^{1/2}_{rr} Q \right ) \sim 4 F_s
\nabla s_0 \, .
\end{equation}
The R.H.S. of the approximate entropy
flux equation, eq.~(\ref{eq:transporteqdriving2}), is a sum of two
important terms: buoyancy driving, $\rho_0 \eta g Q$, and the pressure
covariance, $-\langle
s^{\prime} \nabla P^{\prime} \rangle$.  \citet{launder02} suggest that
this term is best modeled by $-\langle s^{\prime} \nabla P^{\prime}
\rangle \sim -(1/2) \rho_0 \eta g Q$, and our 2D simulations confirm this model.
In effect, the sum of these two source terms reduces buoyancy
driving by a factor of two.

If we approximate the divergence of a generic flux, $F_i$, by
$\nabla_r \cdot F_i \propto - F_i / z$, where $z= R_s - r $ is the
downward distance from the shock, then algebraic solutions of the
approximate equations are proportional to $z^2$.  To obtain the
correct proportionality constants, we assume solutions of the form
$R_{rr} = a z^2$, $F_s = b z^2$, and $Q = c z^2$, and substitute these into
eqs.~(\ref{eq:transporteqdriving1}-\ref{eq:transporteqdriving3}).  This
results in the following algebraic expressions for the second moment correlations
\begin{equation}
\label{eq:kineticalg}
R_{rr} = \frac{2}{9} \eta g (-\partial_r s_0) z^2 \, ,
\end{equation}
\begin{equation}
\label{eq:entropyfluxalg}
F_s = \frac{8^{1/2}}{9}
\rho_0 \left ( \eta g \right )^{1/2} \left ( - \partial_r s_0 \right
)^{3/2} z^2 \, ,
\end{equation}
and
\begin{equation}
\label{eq:entropyvariancealg}
Q = \frac{8}{9}
\left ( - \partial_r s_0 z \right )^2 \, .
\end{equation}
For the entropy gradient in these equations, we assume that
$\partial_r s_0 \approx \dot{Q}/(\dot{m}T_0)$, and to get the radial
component of Reynolds stress we assume $R_{rr} \approx R/3$

If instead of assuming a balance between local driving and transport,
we assume a balance between local driving and dissipation, then we
obtain dimensionally similar results.  However, they differ in the
constants of proportionality and the length scale involved.  With
dissipation, the natural length scale is $\mathcal{L}$, which is a
fixed scale and is either the full
size of the convective region (i.e. a constant) or is proportional to
the local scale height.  In either case, using a fixed length scale
results in algebraic expressions that are inconsistent with our 2D
simulations.  Instead, in \S \ref{section:comparemodels}, we show that the
convective profiles in the driving region of 2D simulations are quite consistent
with $z$ as the length scale.  Interestingly, most contemporary
formulations of MLT use $\mathcal{L}$ as the appropriate length scale,
while in the original formulation of the MLT, Prandtl used $z$.  In
essence, the result that the downward distance, $z$, is the most
appropriate length scale suggests that the properties of
neutrino-driven convection are best modeled by negatively buoyant
plumes that form at the shock and grow due to buoyancy.

\subsection{Model 4: A General Global Turbulence Model}
\label{section:model4}

While the previous models are adequate in some locations, we show in
\S~\ref{section:comparemodels} that they fail to reproduce the global
properties of neutrino-driven turbulence.  Motivated by this failure, we propose
an original turbulence model which is derived using global considerations.

Models 2 \& 3 employ single-point closure models, which assume that on small
scales, the higher order correlations can be modeled using local
2$^{\rm{nd}}$-order correlations. Consequently, some of the most important terms in a nonlocal problem are modeled
using local approximations.  While these local models are adequate in
some locations, they can be factors off in other locations.  Given
the stiff nature of the governing equations, these modest errors can
lead to significantly flawed turbulent profiles.   Rather than relying
on these local models, we use conservation laws and develop an original global turbulence model.

Integrating the turbulence equations leads to global conservation laws
for turbulence.  For example, integrating the
turbulent kinetic energy equation dictates that global buoyant driving
equals global dissipation.  
To satisfy these global constraints, the turbulent correlations relax
into the appropriate profiles.  If the same driving, redistribution, and
dissipation mechanisms operate in a wide range of conditions, then
these global constraints also imply self-similar convection profiles.
For Model 4, we adopt characteristic profiles for the convective
entropy luminosity and turbulent dissipation.  The specific profiles
are motivated by the evolution of large scale buoyant plumes and
informed by numerical 2D CCSN simulations (this paper) and 3D stellar
convection \citep{meakin10}.  Then the scales of these profiles are
constrained using the conservation laws.  Given these scaled profiles, we
then calculate the remaining quantities of interest such as the kinetic
energy flux and entropy profile.  The latter is particularly important
in exploring the conditions for successful CCSN explosions.

Our global model builds upon and generalizes the ideas put forth by \citet{meakin10}\footnote{These global models are similar to the model
  used for the boundary layer in Earth's atmosphere\citep{plate97}}.
In the context of stellar evolution and no background flow,
\citet{meakin10} suggest that the primary role of the entropy
(enthalpy) flux is to redistribute the entropy such that the entropy
gradient is flat and the entropy generation rate is constant
throughout.  In other words, they found universal profiles for the entropy
generation rate and entropy profiles.  They then integrated the
entropy and turbulent kinetic energy equations to provide integral
constraints and solutions for the turbulent scales.

Similarly, the rate of entropy change in the CCSN context is uniform,
in fact during the steady-state phase it is zero everywhere.  However,
the entropy gradient is nonzero, so the \citet{meakin10} model will
not suffice.  2D simulations (see
\S~\ref{section:comparemodels}) indicate that the global constraints
and redistribution drive the convective entropy luminosity, $L_s =
F_s 4 \pi r^2$, and turbulent dissipation to simple self-similar profiles.

Informed by 2D simulations of CCSNe (\S~\ref{section:comparemodels},
Fig.~\ref{mdotds_plot}) and 3D simulations of stellar convection
\citep{meakin10}, we model the convective entropy luminosity with a
piecewise linear function,
\begin{equation}
\label{eq:model4ls}
L_s = \left \{
\begin{array}{lr}
L_{s,0} z/z_0\, , & {\rm for} \, z < z_0\\
L_{s,0} (Z_c-z)/(Z_c-z_0)\, , & {\rm for} \, z > z_0
\end{array}
\right . \, ,
\end{equation}
and the turbulent dissipation via
\begin{equation}
\label{eq:model4dissipation}
\rho \epsilon 4 \pi r^2 \sim \alpha (R_s - r) \, ,
\end{equation}
where $z = r-R_{\ell}$ is the distance from the bottom of the
convective region, $Z_c = R_s - R_{\ell}$ is the total size of the
convective region, and $R_{\ell}$ is defined by
$\int^{R_S}_{R_{\ell}} \dot{Q}/T_0 \, dV = 0$.  In these models, the peak of the
$L_s$ profile is at $z_0$, and goes to zero at $R_{\ell}$ and the
shock.  The turbulent dissipation is zero at the shock, and
increases linearly downward with a scale of $\alpha$.  Given the many
differences (i.e dimensionality, accretion, etc.) between 2D CCSN
and 3D stellar convection simulations, the universality of the profiles
is remarkable and is a testament to their self-similarity.  Taken
together, this model for $L_s$ and $\epsilon$ has three parameters,
$L_{s,0}$, $z_0$, and $\alpha$, which set the scale for convection.

Next, we use the algebraic expression for the entropy flux and global
conservation laws to evaluate these scales.  Operating under the assumption that
the growth of negatively buoyant plumes sets the scale $L_{s,0}$, we
evaluate the algebraic expression for the entropy flux, eq.~(\ref{eq:entropyfluxalg}), at the
peak:
\begin{equation}
\label{eq:model4alg}
L_{s,0} = 4 \pi r^2 F_{s,\textrm{alg}}(z_0) \, .
\end{equation}

We explore two techniques to constrain the position of the peak, $z_0$.  In
the first, we assume that the peak of $L_s$ corresponds to where
entrainment starts to dominate the evolution of negatively buoyant
plumes and that this distance is proportional to the pressure scale
height (i.e. $Z_c - z_0 = \alpha_{z_0} H_p$).  Comparing to the 2D
simulation, we find that $\alpha_{z_0} \simeq 1.7$, 2.1, and 2.0 at 404
ms, 518 ms, and 632 ms after bounce.  Empirically, $\alpha_{z_0} \sim 2$ with
an accuracy of $\sim$10\%.  While this empirical approach provides a
useful diagnostic of $z_0$, it lacks a physical derivation.  

In the second technique, we derive a physically motivated global
constraint for $z_0$.  Consider the approximate entropy flux and
entropy variance equations,
eqs.~(\ref{eq:transporteqdriving2} \& \ref{eq:transporteqdriving3}).
They represent a set of conservation laws for $F_s$ and $Q$ in which
the source terms are determined by buoyant driving.  Combining these
two equations and integrating over the whole convective region, we find that $\int F_s
\cdot \nabla s_0 dr \sim 0$.  In effect, $F_s \cdot \nabla s_0$
represents an important buoyant source term for turbulence, where $F_s
\cdot \nabla s_0$ is negative for the growth (decay) of negatively
(positively) buoyant plumes and vice versa.  Because $F_s = 0$ and
$Q=0$ at both boundaries, the integral of this buoyant source term must
balance out.  Therefore, we adjust $z_0$ so that $F_s$ satisfies
\begin{equation}
\label{eq:model4constraint2}
\int^{R_s}_{R_{\ell}} F_s \cdot \nabla s_0 dr = 0 \, .
\end{equation}

To set $\alpha$, the turbulent dissipation scale, we first rewrite the
turbulent kinetic energy equation as an expression for the turbulent flux:
\begin{equation}
\label{eq:model4kinetic}
\frac{\partial L_k}{\partial r} = \eta L_s g_r - \rho \epsilon 4 \pi
r^2 \, ,
\end{equation}
where we have neglected the term due to background advection.
Integrating eq.~(\ref{eq:model4kinetic}) over the
entire convective region and noting that $L_K$ is zero at the
boundaries leads to the second conservation law for the balance
of total buoyant work and dissipation:
\begin{equation}
\label{eq:model4constraint1}
\int^{R_s}_{R_{\ell}} \eta L_s g_r \, dr = 
\int^{R_s}_{R_{\ell}} \rho \epsilon 4 \pi r^2 \, dr \, .
\end{equation}
Together, eqs.~(\ref{eq:model4alg}, \ref{eq:model4constraint2}, \&
\ref{eq:model4constraint1}) constrain the three scales of our global model.

Having set the scales of the turbulent profiles,
eqs.~(\ref{eq:model4ls}) \& (\ref{eq:model4dissipation}), it is now
possible to integrate eq.~(\ref{eq:model4kinetic}) to find the turbulent
kinetic flux and, most importantly, evaluate the entropy profile
including the effects of turbulence.

Including the time rate of change, the entropy equation is
\begin{equation}
\label{eq:fullentropy}
\frac{\partial (\rho s_0)}{\partial t} + \dot{m} \cdot \nabla s_0 = \frac{\dot{Q} + \rho
  \epsilon}{T_0}-\nabla \cdot F_s \, ,
\end{equation}
and integrating this equation over the volume from an arbitrary
radius, $r$ up to the shock, $R_s$ gives
\begin{equation}
\frac{\partial}{\partial t} \left ( \int^{R_s}_{r}\rho s_0 \, dV
\right ) 
+ \dot{M} \Delta s_0(r) =
L_s(r) + \int^{R_s}_r \frac{\dot{Q}}{T_0} \, dV
+ \int^{R_s}_r \frac{\rho \epsilon}{T_0} \, dV \, ,
\end{equation}
where $\Delta s_0(r) = s_0(R_s) - s_0(r)$ and $L_s(r) = 4 \pi r^2
F_s(r)$.  Assuming a flat gradient, $\Delta s_0$ = 0, and $\partial s_0/\partial t =
\langle (\dot{q} + \epsilon)/T_0 \rangle_m$, where $\langle \rangle_m$
is a mass average, leads to the integral equation that
\citet{meakin10} use for stellar convection.  For the CCSN problem,
we assume steady state, $\partial s_0/\partial t = 0$ and $\Delta s_0
\neq 0$, resulting in
\begin{equation}
\label{eq:model4entropy}
\dot{M} \Delta s_0(r) =
L_s(r) + \int^{R_s}_r \frac{\dot{Q}}{T_0} \, dV
+ \int^{R_s}_r \frac{\rho \epsilon}{T_0} \, dV \, .
\end{equation}

If we momentarily neglect the buoyant term in the kinetic energy
equation, eq.~(\ref{eq:model4kinetic}),
then the profile for the dissipation implies $\partial
L_K/\partial r \sim -\alpha (R_s -r)$, which is reminiscent of the
entrainment hypothesis for the evolution of isolated buoyant plumes
\citep{turner73,rieutord95}.  In Kolmogorov's hypothesis, dissipation
is assumed to be dominated by mechanisms at the largest scale.
Therefore, eq.~(\ref{eq:model4dissipation}) implies that entrainment
of negatively buoyant plumes is {\it the} mechanism at the largest
scales that controls dissipation.

In summary, we use self-similar
profiles of $L_s$ and $\epsilon$ and global conservation laws in Model 4.
Assuming that the growth of negatively buoyant plumes sets the scale
for $L_s$, we evaluate the algebraic model, eq.~(\ref{eq:entropyfluxalg}), at the
peak ($z_0$) of $L_s$.  To determine the location of the peak, we set
$z_0$ such that $L_s$ and $\nabla s_0$ satisfy the global constraint
in eq.~(\ref{eq:model4constraint2}).  Next, the
dissipation scale, $\alpha$, is determined by satisfying the balance
between total buoyant work and dissipation,
eq.~(\ref{eq:model4constraint1}).  Having used global constraints to set
the parameters of $L_s$ and $\epsilon$, we next evaluate the turbulent kinetic luminosity and entropy profiles using
eqs.~(\ref{eq:model4kinetic}) \& (\ref{eq:model4entropy}).  Finally,
we find $R_{rr}$ by inverting our plume model for $F_K$,
eq.~(\ref{eq:fkmodel}).

\section{Comparing Turbulence Models to 2D simulations}
\label{section:comparemodels}

\par In this section, we critically compare the results of 2D
simulations with the turbulence closure models presented in
\S\ref{section:models}.  
Of all the turbulence models that we
explore, Model 4, the global model (\S~\ref{section:model4}),
consistently gives the correct scale, profile, and temporal evolution
for the Reynolds stress, $R_{rr}$, and entropy flux, $F_s$.

\par
\begin{figure}[t]
\epssclone
\plotone{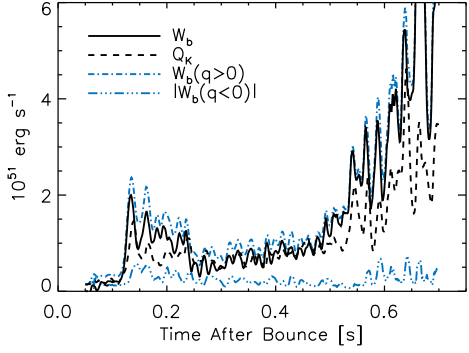}
\caption{Buoyancy work
  integral, $W_b$, (solid) and total turbulent dissipation, $Q_K$
  (dashed), as a function of time after bounce. At all times, the total buoyancy
  work is roughly balanced by the total turbulent dissipation.
  $W_b(q<0)$ (blue, dot-dashed line) is the buoyancy work for the
  region where the
  integrand of $W_b$, $q =
  \langle \rho^{\prime} v_r^{\prime} \rangle g_r$, is greater than
  zero, and $|W_b(q<0)|$ is buoyancy breaking where $q<0$.  At early
  times, buoyancy breaking contributes significantly to the balance, but
  is much diminished as the convective motions strengthen near explosion.
See \S~\ref{section:comparemodels} for further discussion.  \label{convectionintegral}}
\epsscale{1.0}
\end{figure}

 {\em Dissipation.---}  In Fig. \ref{convectionintegral} we present the time history of integrated buoyancy driving and turbulent dissipation.  The integrated buoyancy work $W_b = \int \left < \rho^{\prime} \mathbi{v}^{\prime} \right > \cdot \mathbi{g} \; dV$ should balance the integrated turbulent dissipation $Q_K = \int \rho_0 \epsilon \; dV$ \citep{chandra1961} in steady state.  This can be shown by integrating the turbulent kinetic energy equation (eq.~[\ref{eq:kinetic}]) over the volume of the turbulent region, and assuming that the work done by turbulent pressure fluctuations and the Reynolds stresses are small, and that the flux of kinetic energy is zero out of the turbulent region, all good approximations for the scenario under investigation.    The volume integrated turbulent  kinetic energy equation then reads



\begin{equation}
\label{eq:wbeqek}
\int \left < \rho^{\prime} \mathbi{v}^{\prime} \right >
\cdot \mathbi{g} \; dV
=
\int \rho_0 \epsilon \; dV,
\end{equation}

\noindent or 

\begin{equation}
W_b = Q_K.
\end{equation}

\par The overall balance between $W_b$ and $Q_K$ presented in Fig.~\ref{convectionintegral} shows that the adopted model expression for the dissipation (eq.~[\ref{eq:dissipation}]) leads to an overall consistency with the evolution equation for the turbulent kinetic energy(eq.~[\ref{eq:kinetic}]),  and the simulation data.

\par The global balance between buoyancy driving and turbulent kinetic energy dissipation
tempts one to conclude that all buoyancy work is dissipated by
turbulent dissipation.  While this is true in the global sense that
the net buoyancy work is balanced by turbulent kinetic energy dissipation, it is important
to note that the total buoyancy work is an integral over the turbulent
convection zone of $q = \left < \rho^{\prime} v_r^{\prime} \right >
g_r$, which is positive in the active driving region and negative in
the bounding stabilizing layer.  Therefore, some of the work done by
buoyancy in the active {\em driving} region ($q>0$) is mitigated by
the {\em buoyancy breaking} ($q<0$) that takes place in the
stabilizing layer.  It is therefore only the net buoyancy work that is
balanced by the total turbulent dissipation.  During the earliest
stages buoyancy breaking is significant and is about half the
magnitude of dissipation.  As convection builds in strength after 250
ms, the significance of buoyancy breaking steadily diminishes until it
is about 1/10 of dissipation.

\par
\begin{figure}[t]
\epsscltwo
\plotone{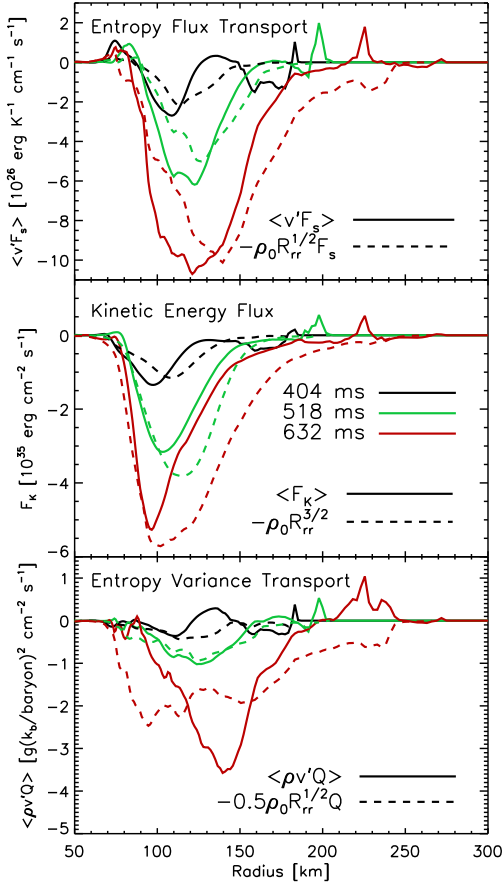}
\caption{Comparison of turbulent transport terms (solid lines) to model
  eqs.~(\ref{eq:entropyfluxtransportmodel}-\ref{eq:entropyvariancetransportmodel})
  (dashed lines).
  This figure is similar to Fig. \ref{transport3times} in
  presentation.  The transport terms are not
  proportional to the gradient of $R_{rr}$ as the gradient
  approximation would suggest \citep{pope00,launder02}.  Rather, they
  are proportional to $R_{rr}^{1/2}E_i$, where $E_i$ is the quantity in
  question to be transported.  This is a consequence of
  transport by large scale plumes.  \label{transportmodels}}
\epsscale{1.0}
\end{figure}

 {\em Turbulent  Fluxes. ---} In Fig.~\ref{transportmodels} we
present profiles of the turbulent flux for three quantities, entropy
flux, turbulent kinetic energy, and entropy variance.  A comparison to
the gradient diffusion approximation confirms earlier results that it
is a poor model for transport in thermal convection.  Instead, we find
relatively good agreement with the transport models proposed in \S
\ref{section:transportmodels} which are proportional to
${R_{rr}}^{1/2} E_i$, where $E_i$ is the density of the transported
quantity.  The agreement between these models and the data confirm the
advective nature of the transport (i.e. $F_i \propto v' E_i$).

\subsection{Model Comparisons for $\langle Q\rangle$, $\langle R_{rr}\rangle$, and $T_0 \langle F_s\rangle$}

\par
\begin{figure}[t]
\epsscltwo
\plotone{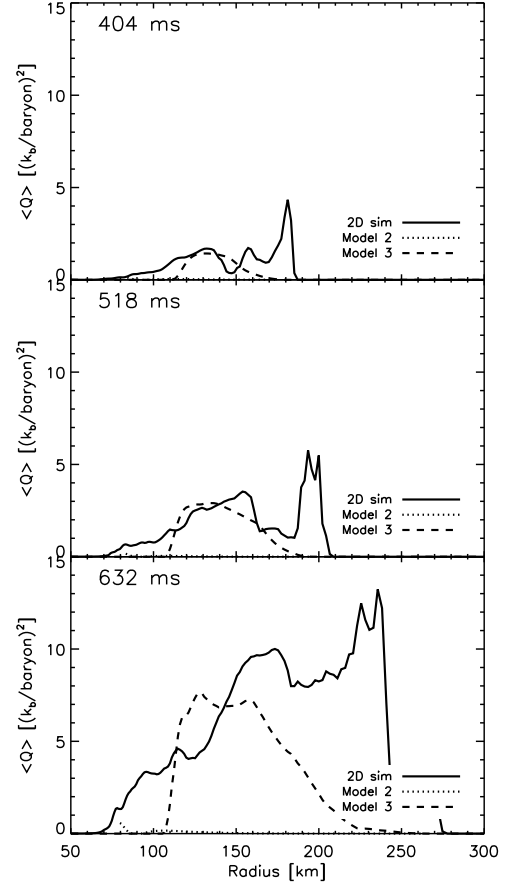}
\caption{Radial profile of the
  entropy variance, $\left < Q \right > $ during the three
  phases shown in Fig. \ref{convectionsasistills}.  In this figure, we
  compare the results of 2D simulations (solid line) with the results
  of the turbulence models.  Only Models 2 \& 3 (\S
  \ref{section:model2} \& \ref{section:algebraic}) provide any
  predictions for $\left < Q \right >$, and of these the algebraic
  model (dashed-line, Model 2) gives the appropriate scale and
  evolution.  \label{compsimmodels_sp2}}
\epsscale{1.0}
\end{figure}

\begin{figure}[t]
\epsscltwo
\plotone{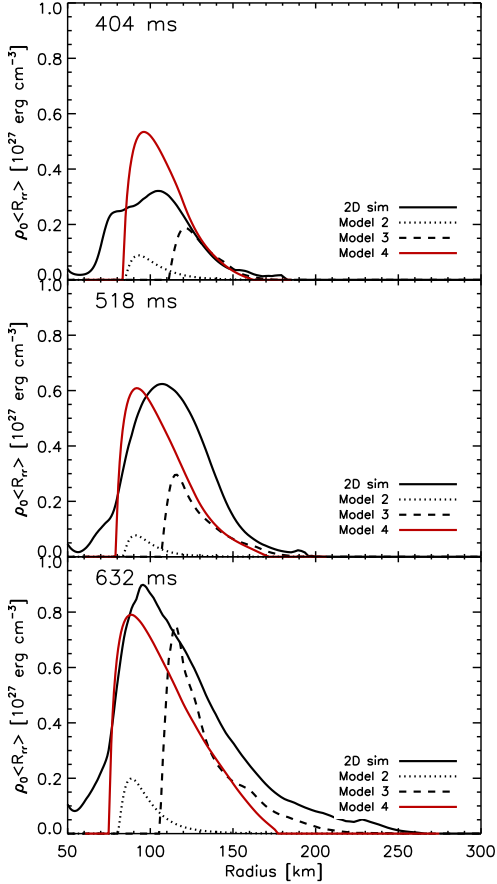}
\caption{Radial profile of the
  radial part of the Reynolds stress, $\left < R_{rr} \right >$ during
  the three phases shown in Fig. \ref{convectionsasistills}.  As in
  Fig. \ref{compsimmodels_sp2}, this figure compares the results of 2D
  simulations (solid lines) with the results of the models presented
  in sections
  \ref{section:model2}-\ref{section:model4}.  The 2D results
  best match the radial profile and the temporal evolution of our
  global model (red line, Model 4).
\label{compsimmodels_rey}}
\epsscale{1.0}
\end{figure}

\begin{figure}[t]
\epsscltwo
\plotone{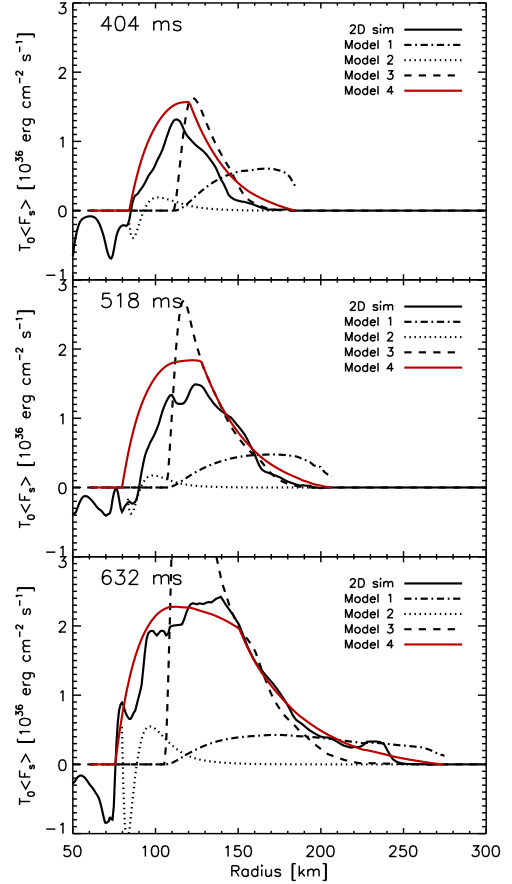}
\caption{Radial profile of the
  enthalpy flux, $T_0 \left < F_{s} \right >$ during the three phases
  shown in Fig. \ref{convectionsasistills}.  This figure compares the
  results of 2D simulations (solid lines) with the results of the
  models presented in sections
  \ref{section:model1}-\ref{section:model4}.  Of these models, only
  the global model (red line, Model 4) reproduces the scale, profile, and evolution of the
  2D simulation. \label{compsimmodels_fs}}
\epsscale{1.0}
\end{figure}

\begin{figure}[t]
\epsscltwo
\plotone{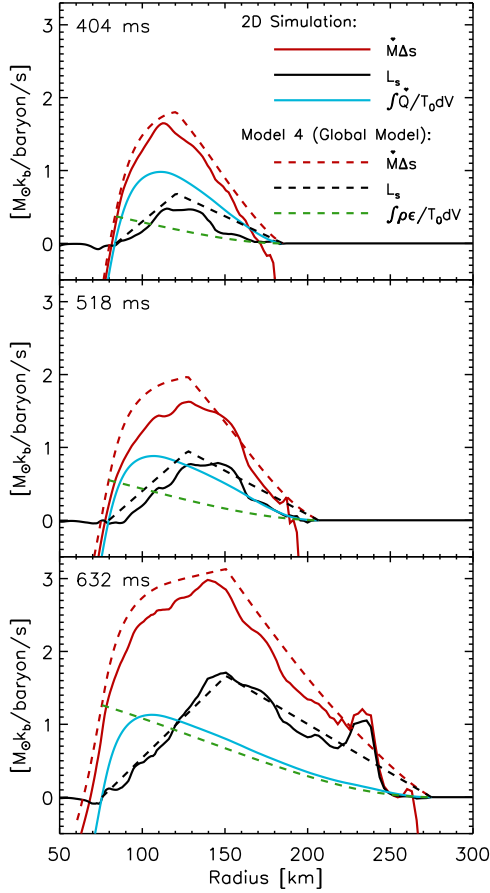}
\caption{Radial
  profiles of entropy change in units of $\dot{M} \Delta s$. The solid
  lines represent 2D simulation results, and the dashed lines are the
  results of the global model, Model 4.  The
  integrated neutrino heating and cooling (blue line) and turbulent
  dissipation (green dot-dashed line) are evaluated from $r$ to the
  shock radius, $R_s$.  The modeled entropy difference (red-dashed line) is a sum of the integrated
  neutrino heating and cooling (blue line), the modeled integrated
  turbulent dissipation (green-dashed line), and the modeled
  convective entropy luminosity, $L_s$ (green dot-dashed line).The
  entropy difference derived using the global model (red-dashed line) shows the same scale, radial profile, and temporal
  evolution as the 2D simulation data (red solid line).  
  Convection ($L_s > 0$) fills the region where the integrated
  neutrino heating and cooling (blue line) is greater than zero.
  The 2D simulated $L_s$ profile (black-dashed line) is self-similar
  and can be modeled by a piecewise linear, pointed
  hat. \label{mdotds_plot}}
\epsscale{1.0}
\end{figure}

\begin{figure}[t]
\epsscltwo
\plotone{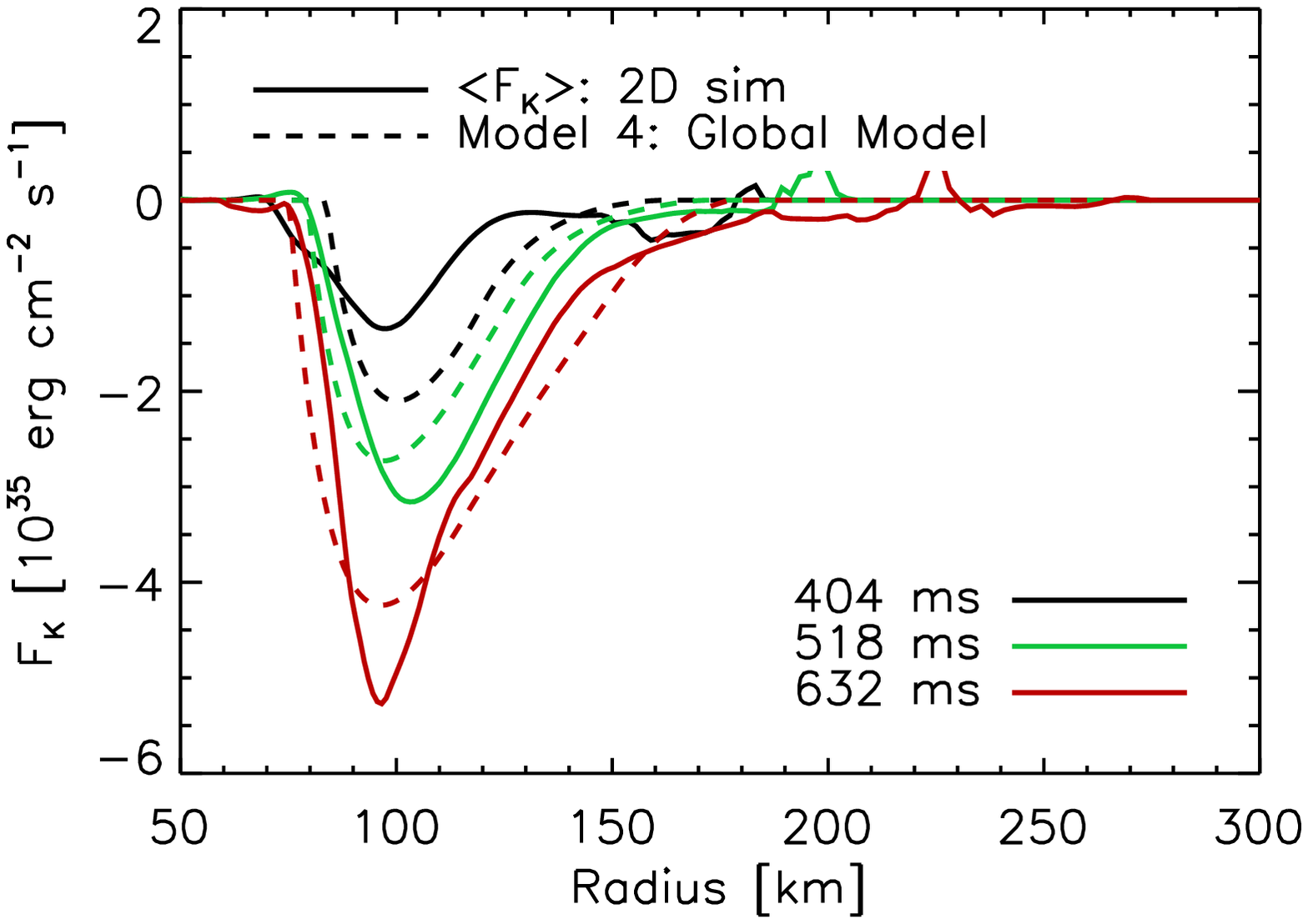}
\caption{Convective kinetic energy flux, $F_K$, at 404
    ms, 518 ms, and 632 ms after bounce as a function of radius.  The
    results of the global model (Model 4, dashed lines) reproduce the
    scale, general profile, and temporal evolution of the 2D data
    (solid lines). \label{compsimmodels_fk}}
\epsscale{1.0}
\end{figure}

 In this subsection we critically compare each of the turbulence
models introduced in \S\ref{section:models} to the simulation
data. This comparison is summarized by Figures
\ref{compsimmodels_sp2}-\ref{compsimmodels_fk}.  The first three show
the radial profiles of $\left < Q \right >$, $\left < R_{rr} \right
>$, and $T_0 \left < F_s \right >$ for the 2D simulation data and the
turbulence models of \S~\ref{section:model1}-\ref{section:algebraic}.
All three figures are divided into three panels with each panel
showing the 2D simulation profile (solid line) and the turbulence
models at the three times of Fig.~\ref{convectionsasistills}.
The
last two figures, Figs. \ref{mdotds_plot} \& \ref{compsimmodels_fk},
compare the entropy, entropy flux, and kinetic energy flux profiles of
the 2D simulations with the profiles produced by our global model
(Model 4, \S~\ref{section:model4}).

\par {\em Model 1 comparison (\S\ref{section:model1}).---} The zero
  entropy gradient (Model 1, \S~\ref{section:model1}) model is shown
  as dot-dashed lines in
  Figs.~\ref{compsimmodels_sp2}-\ref{compsimmodels_fs}. Though this
  model gives the correct order of magnitude for $\left < T_0 F_s
  \right >$, it fails to reproduce the specific radial profile and
  temporal evolution.  In fact, there appears to be no temporal
  evolution in this turbulence model, while the 2D results clearly
  grow with time.  Furthermore, the zero entropy gradient model (Model
  1) gives a non-zero entropy flux at the shock. This non-zero entropy
  flux is clearly not a characteristic of the 2D simulations.  See
  \S~\ref{section:model1} for a discussion of how this non-zero
  entropy flux corrupts the solutions for the steady-state accretion
  shock.

\par {\em Model 2 comparison (\S\ref{section:model2}).---} The Reynolds stress and heat flux closure model is represented by a dotted-line in each plot.  Of all of the turbulence models presented in this paper, it produces the poorest reproduction of the turbulence profiles. To obtain these profiles we used a shooting method to integrate eqs.~(\ref{eq:reynoldsstressmodel}-\ref{eq:entropyvariancemodel}) subject to the background flow and boundary conditions.  At present it is not clear what the specific form for the boundary conditions should be, especially at the shock, so we used the values of $\left <   Q \right >$, $\left < R_{rr}\right > $, and $\left <  F_s \right > $ at a small distance from the shock.  We then integrated eqs.~(\ref{eq:reynoldsstressmodel}-\ref{eq:entropyvariancemodel}) inward until $\left < R_{rr} \right > = 0$ at the inner boundary.  We adjusted the guess for $F_s$ at the outer boundary so that both $F_s$ and $R_{rr}$ are zero at the lower boundary.  

\par For several reasons, we strongly disfavor this model.  The most obvious reason is the lack of consistency between the model and the 2D results.  In addition, because these equations are stiff, they are quite sensitive to the boundary conditions and the assumptions for dissipation.  We adjusted the distance where we sampled the boundary conditions just below shock and found the solutions to be extremely sensitive to this location.  Others have noted similar convergence problems with boundary conditions to these equations \citep{wilcox06}.  Finally, this model has many parameters that have been calibrated for the solution of isolated buoyant plumes, not for fully developed convection.

\par {\em Model 3 comparison (\S\ref{section:algebraic}). ---}  In the
region where convection is being actively driven, the algebraic model,
eqs.~(\ref{eq:kineticalg}-\ref{eq:entropyvariancealg}), produces
reasonably accurate profiles and temporal evolution.  Of the three
turbulent moments shown in
Figs. \ref{compsimmodels_sp2}-\ref{compsimmodels_fs}, $R_{rr}$ and
$T_0 F_s$ matter most in the background equations,
eqs.~(\ref{eq:mass}-\ref{eq:entropy}), and they show the best
correlation with 2D simulations.  On the other hand, while the algebraic
model gives the correct scale for the entropy variance, $Q$, the
algebraic model does not match exactly
the 2D profiles.  Fortunately, the entropy variance does not directly
influence the background equations and so in practice this discrepancy
can be ignored. Nonetheless, this failure should be a clue to what is
missing. Furthermore, we note that the profiles for $R_{rr}$ and
$T_0F_s$ match the 2D results only in the heating region, where
convection is actively driven.  Below the heating region ($r \lesssim
100$km), we set the values to zero because it is not clear how to
model this region with the algebraic model where positive buoyancy
decelerates the convective plumes.  Finally, as we discuss in
\S~\ref{section:algebraic}, the success of this model implies that
core-collapse turbulence is best characterized by low entropy plumes
that are initiated at the shock, the acceleration of these plumes
through the heating region, and the deceleration of these plumes at
the lower boundary by stabilizing gradients.

\par {\em Model 4 comparison (\S\ref{section:model4}). ---}  Figures
\ref{compsimmodels_rey}-\ref{compsimmodels_fk} show that the global
model (Model 4) provides the most accurate turbulent correlations.
The Reynolds stress, $R_{rr}$, and enthalpy flux, $T_0 F_s$, (red
solid lines in Figs.~\ref{compsimmodels_rey} \&
\ref{compsimmodels_fs}) derived from Model 4 have profiles that match
the 2D simulation data in scale, shape, and temporal evolution.

Figure \ref{mdotds_plot} compares the terms in the entropy equation,
eq.~(\ref{eq:model4entropy}), of Model 4 with 2D simulation data, and once again, this plot
shows that Model 4 accurately reproduces the 2D data.  The solid blue
line represents the change in entropy (in units of $\dot{M} \Delta
s_0$) due to neutrino heating and cooling alone.  We find that
convection fills the region where this integral is greater than zero.
In the convective region, the neutrino heating and cooling curve
accounts for only half of the total entropy change at 404 ms and
only one third of the entropy change at 632 ms.  Heating by turbulent
dissipation and redistribution by $L_s$ account for the rest.  The total
entropy change, $\dot{M} \Delta s_0$, from Model 4 (red-dashed line)
is computed by summing the neutrino heating and cooling integral (blue
line), the modeled turbulent dissipation integral (green-dashed line), and the modeled convective entropy luminosity (black-dashed
line).  The modeled entropy difference is only slightly larger than
the 2D simulation results (red-solid line) and reproduces the
general radial profile and temporal evolution.  

In \S~\ref{section:model4}, we argue that the global constraints of
convection and similarity in driving, distribution, and dissipation
mechanisms suggest self-similar profiles for $L_s$.  Indeed, the
correspondence between our modeled $L_s$ (green dot-dashed line) and the 2D
data (black dashed line) confirms this assumption.  Moreover, this
shape is simply modeled as a piecewise linear, pointed hat.  The scale
of $L_s$ is set by the entropy flux we derive from the algebraic model
(Model 3, \S~\ref{section:algebraic}) at the position of the peak.  Since the
algebraic model describes the growth of negatively buoyant plumes that
originate at the shock, the scale of $L_s$ is in turn set by the
growth of these negatively buoyant plumes.  The position of the peak
is determined such that the integral constraint, $\int F_s \cdot
\nabla s_0 \, dr = 0$, eq.~(\ref{eq:model4constraint2}), is satisfied.

In Fig.~\ref{compsimmodels_fk}, we compare the kinetic energy
flux of Model 4 (dashed lines) with the results of 2D simulations
(solid lines).  Qualitatively, the modeled fluxes exhibit the correct
scales, radial profiles and temporal evolution.

\section{Turbulence and Conditions for Explosion}
\label{section:conditions}


In this section, we suggest that the entropy equation,
eq.~(\ref{eq:entropy}), holds the key to understanding the explosion conditions.  Furthermore, we use this
equation to argue that of all the convective terms, the divergence of
the convective entropy flux most affects the critical luminosity for successful
explosions.  It has been suggested that the critical luminosity
condition is equivalent to either a ratio of timescales condition
\citep{thompson00,janka01,thompson05,murphy08b} or
to an ante-sonic condition \citep{pejcha11}.  In either case, it is the entropy
equation, eq.~(\ref{eq:entropy}), that leads to this result.  For
example, if we ignore the turbulence terms and
integrate the entropy equation over the gain region, then we can
derive a ratio of advection to heating timescales
\begin{equation}
\label{eq:timecondition}
\frac{\tau_{\rm{advect}}}{\tau_{\rm{heat}}} \sim \frac{1}{\Delta s}
\int^{r_{\rm{gain}}}_{R_s} \frac{\dot{Q}}{\dot{m} T} dr \, ,
\end{equation}
where $\Delta s = s(r_{\rm gain}) - s(R_s)$ is the change in the
entropy between the shock ($R_s$) and gain ($r_{\rm{gain}}$) radii.
Numerical results have confirmed that explosion ensues when this ratio
exceeds $\sim$1 \citep{buras06a,scheck08,murphy08b,nordhaus10}.  In the simplest
interpretation, this result suggests that explosion occurs roughly when
$\Delta s(r)$ exceeds a critical value, $\Delta s_{\rm{crit}}$.  Including turbulent
terms, the change in entropy as function of radius is
\begin{equation}
\label{eq:newtimecondition}
\Delta s(r) = 
-\int^{R_s}_{r} \frac{\dot{Q}+\rho_0 \epsilon}{\dot{m} T} dr
- \frac{L_s(r)}{\dot{M}} \, ,
\end{equation}
where $L_s(r) = 4 \pi r^2 F_s(r)$ and recall that $\dot{m}$ and
$\dot{M}$ are negative. In our 2D simulations, we
find that, of the two turbulent terms in
eq.~(\ref{eq:newtimecondition}), the last term contributes the most
entropy change.  Therefore, if the time condition is a relevant
explosion condition, then the entropy flux
 is the turbulent correlation that most affects the critical
 luminosity.  

While the timescale condition has proven to be a useful diagnostic for
explosions, \citet{pejcha11} have found a more precise explosion
condition in that explosions occur when $c_s^2/v_e^2$ exceeds $\sim$
0.2, where $c_s^2$ and $v_e^2$ are the local sound speed and escape
velocities squared.
They call this new condition the ante-sonic condition and note that it
varies by only a few percent when the neutrino luminosity is changed
by two orders of magnitude.  Over the same range of
neutrino luminosities, the timescale ratio at explosion varies from
0.7 to 1.1.  In other words, the timescale condition is of order 1, but it varies by
$\sim$1.6.  Hence, while the timescale ratio is a useful diagnostic
relating the most important physical processes, the ante-sonic
condition is a more precise (although more obscure) condition for explosion.

Using the integrated form of the entropy equation, eq.~(\ref{eq:newtimecondition}), we show
that the timescale diagnostic and ante-sonic condition are intimately related.  The difference in the sound speed
between the shock and an arbitrary radius, $r$ is 
\begin{equation}
\label{eq:soundspeed}
\Delta c_s^2(r) \approx \frac{P}{\rho_0} + \frac{c_T^2}{c_V} \Delta s(r)
\, ,
\end{equation}
where $P$ and $\rho_0$ are evaluated at $r$, $c_T^2 =
-(T/\rho)(\partial P/\partial T)_{\rho}$ is evaluated at constant
density, and $c_V = T (\partial s / \partial T)_V$ is the specific
heat at constant volume.  The first term on the R.~H.~S. of
eq.~(\ref{eq:soundspeed}) gives the increase in
sound speed due to adiabatic compression.  The second term represents
the change in the sound speed due to changes in entropy given by heating and cooling,
\begin{equation}
-\frac{c_T^2}{c_v}\int^{R_s}_{r} \frac{\dot{Q}}{\dot{m} T} dr \, ,
\end{equation}
and by convection,
\begin{equation}
- \frac{c_T^2}{c_v}\frac{L_s}{\dot{M}} \, .
\end{equation}
Furthermore, since the second term is
proportional to the change in entropy, it is also proportional to
the ratio of timescales.  
Through $\Delta s$, it is apparent that the ante-sonic condition,
$c_s^2/v_e^2 \sim 0.2$, is directly related to the timescale
diagnostic.

In a forthcoming paper, we will
provide a more thorough discussion of these conditions, how they
relate to the critical luminosity for explosions, and how convection
affects all three conditions.  For now, these analytics reaffirm the
supposition in \S~\ref{section:validateequations} that the convective
entropy flux most affects the explosion conditions.

\subsection{Turbulence and Explosion Condition: 2D Simulation Results}

\begin{figure}[t]
\epsscltwo
\plotone{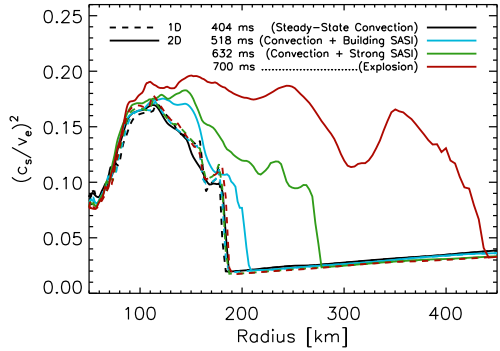}
\caption{Ratio of sound speed and escape velocity squared as a
  function of radius.  The four times shown correspond to the three
  phases in Fig.~\ref{convectionsasistills} (404, 518, and 632 ms
  after bounce) and explosion (700 ms after bounce).  As convection
  increases in vigor, the maximum in this ratio increases until it
  reaches $\sim$0.2 at the time of explosion.  These results are consistent with
  the ante-sonic condition proposed by
  \citet{pejcha11}.  \label{ratiospeeds3times}}
\epsscale{1.0}
\end{figure}

To see if the ante-sonic condition is
consistent with our 2D simulations, we plot $c_s^2/v_e^2$ as a function
of radius and at four times in Fig. \ref{ratiospeeds3times}.  The
first three times correspond to the stages shown in
Fig. \ref{convectionsasistills} and sample a range of convective
strength from weakest at the earliest time to strongest at the latest
time.  The final time, 700 ms after bounce, corresponds to the time of
explosion, which we define as the time when all measures of shock
radii (see the top panel in Fig.~\ref{convectiondata}) expand
indefinitely.  For comparison, we show $c_s^2/v_e^2$ of 1D models for
these times.  The 1D profiles show very little evolution.  However, in
2D simulations, the maximum of $c_s^2/v_e^2$ is strongly correlated
with the strength of convection and the shock radius.  At explosion,
the peak of $c_s^2/v_e^2$ is $\sim$0.2, which is consistent with the
explosion condition proposed by \citet{pejcha11}.

It has been argued that convection increases the dwell time in the
gain region, which in turn reduces the critical luminosity.
\citet{pejcha11}, on the other hand, propose that convection acts to
rearrange the flow so that there is less cooling, and this reduced
cooling is responsible for lower critical luminosities.  The heating
and cooling profiles in Fig. \ref{heating3times} and the entropy
profiles in Fig. \ref{entropy3times} offer a way to investigate the
merits of each proposal.  Unfortunately, interpretation is
somewhat complicated by the fact that the average 2D cooling is less
than 1D cooling for some radii and times but is higher for other radii
and times.  Below $\sim$80 km, 2D cooling is always more than 1D
cooling by $\sim$10\%.  Above this radius, 2D cooling is generally a
few percent less than 1D cooling.  However, at later times (518 and
632 ms) and above $\sim$120 km, 2D cooling is a few percent larger than
1D again.

However, Fig. \ref{entropy3times} shows that the differences in the
average cooling between 1D and 2D are small and do not greatly affect the entropy
profile.  When the convective terms are ignored, the 2D (solid green) and 1D (solid
red) entropy profiles are quite similar.  Though there are small differences, the
differences in average cooling profiles are dwarfed by the effects of
including the convective terms in the entropy equation (dot-dashed
green curve).  Therefore, it is unlikely that changes in the average
cooling between 1D and 2D lead to the reduction in the critical
luminosity.  Rather, as we argue in \S \ref{section:conditions} it is
more likely that the divergence of the convective entropy flux is
responsible for the extra entropy, higher sound speeds, and a
reduction in the critical luminosity.

\section{Conclusion}
\label{section:conclusion}

Recent simulations of CCSNe indicate that turbulence reduces the
critical neutrino luminosity for successful explosions.  This
suggests that a theory for successful explosions requires a
theoretical framework for turbulence and its influence on the
critical luminosity.  In this paper, we develop a foundation
for this framework, which is represented by the following results:
\begin{packed_item}
\item We derive the exact steady-state equations for the background and turbulent flow.
\item We identify the convective terms that most influence the
  conditions for successful explosions. 
\item We have shown that without turbulence, entropy profiles of 2D
  simulations would be nearly identical to 1D and that the convective
  terms entirely make up the difference.  
\item This further motivates the need to understand turbulence in the
  context of CCSNe.  To this end, we cull the literature for a broad
  sample of turbulence models, but after a quantitative comparison
  with 2D simulations, we find that none adequately reproduce the
  global turbulent profiles.  These single-point models fail because
  they use local closure approximations, even though buoyantly driven
  turbulence is a global phenomenon.
\item Motivated by the necessity for an alternate approach, we propose
  an original model for turbulence which incorporates global
  properties of the flow.  This global model has no free
    parameters; instead the scale (or parameters) of convection are constrained
    by global conservation laws.  Furthermore, this model accurately
  reproduces the turbulence profiles and evolution of 2D CCSN
  simulations.  
\end{packed_item}

Using Reynolds decomposition, we derive steady-state averaged equations for the
background flow and turbulent correlations,
eqs.~(\ref{eq:reynoldsstress}-\ref{eq:entropyvariance}).  These equations naturally incorporate
effects that are important in the CCSN problem such as
steady-state accretion, neutrino heating and cooling, non-zero entropy
gradients, buoyant driving, turbulent transport, and dissipation.  
We validate these equations using 2D CCSN simulations.  For example, we
integrate the entropy equation with and without the convective terms
(see Fig. \ref{entropy3times}).  If we neglect the turbulence terms, then we
recover the 1D entropy profile.  The difference between the 1D and 2D
entropy profiles is entirely accounted for by the physics of turbulence.

Turbulence
equations require closure models, but these closure models depend upon
the macroscopic properties of the flow.  
To derive a closure model
that is appropriate for CCSNe, we compare a representative sample of closure models in the literature with 2D simulations.
Motivated by the failure of these models,
we have developed an original closure model.  While the models culled
from the literature are single-point closure models and use local
closure approximations, our model is distinguished by using global
properties of the flow for closure.  This global model is
further distinguished by reproducing the scale, profile, and evolution
of turbulence in 2D simulations.

The single-point models use local turbulent correlations to derive closure
relations for the higher order correlations.  Convection is inherently
a global phenomenon, and so while it is possible to model the
higher-order correlations with local approximations in some locations,
these models can be factors off in other locations.
Given the stiff nature of the Reynolds-averaged equations, these
errors, even if modest, can lead to significantly flawed global solutions.   
Rather than relying on these local models, we integrate the turbulence
equations and derive global constraints based on conservation laws.
We propose that nonlocal turbulent
transport relaxes the turbulent profiles to satisfy these global
constraints.  This relaxation combined with the similarity of buoyant
driving, entrainment, and dissipation leads to self-similar profiles
for the most important turbulent correlations.  In Model 4, we
construct a global model in which we define these self-similar
profiles and use global conservation laws to determine their scales.
Locally, we use the differential form of the conservation equations to
derive the remaining profiles. 

Our model represents a new approach to turbulence modeling, so we
elucidate the assumptions and features that distinguish it from
previous models.  Single point closure models try to employ universal
characteristics on the smallest scales to close the problem.  We are
approaching this from the other direction.  The nonlocal nature of
plume dominated convection leads us to assume universality on
the largest scales and a minimum set of global profiles to close the problem.
These two approaches are complimentary.  Assuming universality on the
smallest scales lends itself to dynamic simulations, while the global
approach lends itself very well to steady-state problems.

The general strategy that we employ is to establish some general
characteristic of turbulence and use global conservation laws to
constrain the scale.  For now, we identify the apparent self-similar
profiles as the general characteristic.  In fact, these self-similar
profiles are motivated by the generic properties of plume dominated
flows and the results of 2D CCSN and 3D stellar convection
simulations.  In the future, we hope to identify a more fundamental
characteristic and physical assumption that leads to these profiles.
But until then, our global model is the only model that consistently
gives the correct scale, profile, and temporal evolution for the
convective kinetic energy flux, $F_{K}$, and entropy flux, $F_s$.  The strongest
validation of this model is Fig. \ref{mdotds_plot}, in which we
reproduce the entire entropy profile of 2D simulations.

In preparation to deriving the reduced critical luminosity, we
identify the turbulent terms that most influence the conditions for explosion.
Three explosion conditions have been explored in the literature.
\citet{burrows93} proposed a critical neutrino luminosity for
successful explosions and \citet{murphy08b} used 1D and 2D simulations
to show that this condition indeed separates steady state accretion
from dynamic explosions.  Alternatively, it has been suggested that
explosions occur when the advection timescale through the gain region
exceeds the heating timescale.  More recently, \citet{pejcha11}
suggest an ante-sonic condition in which explosions occur once
$c^2_s/v^2_e$ exceeds 0.2.
In Fig. \ref{ratiospeeds3times}, we show that in 2D simulations,
$c^2_s/v^2_e$ indeed reaches 0.2 at explosion.  Moreover, using our
Reynolds-averaged equations, we show that the timescale and the
ante-sonic conditions are intimately related, and in both conditions,
convection aides explosion because turbulence raises the entropy
by a term proportional to $L_s/\dot{M}$.

In summary, our global turbulence model contains no free parameters,
is globally self-consistent, accurately reproduces the mean-field
properties of 2D CCSN turbulence, and promises to explain the
reduction in the critical luminosity.
Despite these successes, closure approximations generically depend upon
the properties of the macroscopic flow, making them case-dependent.
Hence, it is unclear to what extent this turbulence model can
accurately describe 3D turbulence, especially in the presence of rapid
rotation and/or magnetic fields.  

Preliminary work \citep{thompson05,yamasaki05,burrows07c} hints that
large rotation rates and magnetic field strengths could aid
explosion, but it is uncertain how modest values would alter turbulence
and its effects on explosion.  Under the most extreme rotation
rates and/or magnetic fields, the flow can be severely distorted from
spherical symmetry \citep[e.g. jets,][]{burrows07c}. In these
conditions, it is best to study the role of rotation and magnetic
fields on turbulence using multi-dimensional simulations.  On the other hand, for
mild rotation and magnetic fields, the Reynolds decomposition
framework employed in this paper can be applied straightforwardly:
mild rotational effects can be included by retaining off-diagonal
Reynolds stress terms \citep[e.g.,][]{garaud10} and applying Reynolds
decomposition to ideal MHD introduces terms associated with the
fluctuations of magnetic fields such as Maxwell stresses and Ohmic
heating \citep[e.g.,][]{ogilvie03,pessah06}.  However, these analyses
are beyond the scope of this paper, so for now, we comment on the
reliability of our global turbulence model for 3D CCSN turbulence.

Even though 2D and 3D turbulence are known to behave differently, the
global turbulence model reproduces the turbulent characteristics of 2D
CCSN \emph{and} 3D stellar evolution simulations.  We suspect, but have not
proven, that this is a testament to the global nature of the
turbulence model.  Though encouraging, there is no guarantee that the
model will work so well for 3D CCSN simulations.
Therefore, a reliable turbulence closure model will require comparison
with 3D simulations.  In 2D simulations, steady-state is a valid assumption.
However, differences in the plume structure of 3D turbulence could
lead to more efficient heating, which in turn could necessitate
including time-dependent terms in the turbulence equations.
The global nature of turbulence and the similarity of driving and
dissipation should lead to self-similar profiles in both 2D and 3D
turbulence.  However, the exact profiles may differ.  Whether any of
these differences will affect closure approximations is uncertain.
Only comparison with 3D simulations can clear up this matter.

\section*{Acknowledgments}
We thank Jason Nordhaus and Ondrej Pejcha for their comments on this
manuscript.  J.W.M. is supported by an NSF Astronomy and
Astrophysics Postdoctoral Fellowship under award AST-0802315.  The
work by Meakin was carried out in part under the auspices of the
National Nuclear Security Administration of the U.S. Department of
Energy at Los Alamos National Laboratory and supported by Contract
No. DE-AC52-06NA25396.


\end{document}